%% file: main.tex
\documentclass[a4paper,UKenglish,cleveref, autoref, thm-restate,colorlinks=true]{lipics-v2021}

\pdfoutput=1 
\hideLIPIcs  

\title{Semantic Bounds And Multi Types, Revisited} 

\author{Beniamino Accattoli}{Inria \& LIX, \'Ecole Poytechnique, France}{beniamino.accattoli@inria.fr}{https://orcid.org/0000-0003-4944-9944}{}

\authorrunning{B. Accattoli} 

\Copyright{Beniamino Accattoli} 



\category{} 

\nolinenumbers 

\EventEditors{John Q. Open and Joan R. Access}
\EventNoEds{2}
\EventLongTitle{42nd Conference on Very Important Topics (CVIT 2016)}
\EventShortTitle{CVIT 2016}
\EventAcronym{CVIT}
\EventYear{2016}
\EventDate{December 24--27, 2016}
\EventLocation{Little Whinging, United Kingdom}
\EventLogo{}
\SeriesVolume{42}
\ArticleNo{23}

\input{\macrospath/macros-preamble}
\setboolean{lipicsstyle}{true}
\input{\macrospath/macros-ben-giulio}
\setboolean{techreport}{true}
\begin{document}

\maketitle

\input{00-Abstract}

\input{01-Introduction}
\input{02-Reductions}
\input{03-Types}
\input{04-Bounds_From_Derivations}

\input{05-Bounds_From_Types}
\input{06-Dissecting_Bounds_From_Types}

\input{07-Bounds_From_Composable_Types}
\input{08-Head_Case}

\input{10-Conclusions}

\bibliography{main.bbl}

\techreport{
\clearpage
\appendix
\input{99-00-Appendix}
}

\end{document}

%% file: macros/macros-preamble.tex
\usepackage{ifthen}
\usepackage{xspace}

\newboolean{talk}
\setboolean{talk}{false}
\newboolean{paper}
\setboolean{paper}{false}
\newboolean{techreport}
\setboolean{techreport}{false}

\newboolean{IEEEstyle}
\setboolean{IEEEstyle}{false}
\newboolean{lipicsstyle}
\setboolean{lipicsstyle}{false}
\newboolean{eptcsstyle}
\setboolean{eptcsstyle}{false}
\newboolean{entcsstyle}
\setboolean{entcsstyle}{false}
\newboolean{sigplanstyle}
\setboolean{sigplanstyle}{false}
\newboolean{easychairstyle}
\setboolean{easychairstyle}{false}
\newboolean{scrartclstyle}
\setboolean{scrartclstyle}{false}
\newboolean{lncsstyle}
\setboolean{lncsstyle}{false}

\newboolean{acmmart}
\setboolean{acmmart}{false}

\newboolean{needstheorems}
\setboolean{needstheorems}{false}
\newboolean{withimages}
\setboolean{withimages}{false}
\newboolean{withproofs}
\setboolean{withproofs}{true}

\newboolean{french}
\setboolean{french}{false}

%% file: macros/macros-ben-giulio.tex
\usepackage{ebproof}
\ebproofset{label separation=0.3em}

\usepackage[utf8]{inputenc}
\usepackage{amsmath, amsfonts,	amsthm, mathtools}
\usepackage{stmaryrd}
\usepackage{marginnote}

\usepackage{appendix}
\capturecounter{theorem}
\capturecounter{figure}

\newcommand{\sem}[1]{\llbracket#1\rrbracket}

\newcommand{\ignore}[1]{}

\newcommand{\myinput}[1]{\ifthenelse{\boolean{withimages}}{\input{#1}}{}}

\newcommand{\reflemma}[1]{Lemma~\ref{l:#1}}
\newcommand{\reflemmap}[2]{Lemma~\ref{l:#1}.\ref{p:#1-#2}}

\newcommand{\reflemmaeq}[1]{{L.\ref{l:#1}}}

\newcommand{\refpoint}[1]{Point~\ref{p:#1}}

\newcommand{\refpointeq}[1]{P.\ref{p:#1}}

\newcommand{\refthmeq}[1]{{T.\ref{thm:#1}}}

\newcommand{\refthm}[1]{Thm.~\ref{thm:#1}}
\newcommand{\refthmp}[2]{Thm.~\ref{thm:#1}.\ref{p:#1-#2}}

\newcommand{\refprop}[1]{Prop.~\ref{prop:#1}}
\newcommand{\refpropeq}[1]{Pr.~\ref{prop:#1}}
\newcommand{\refpropp}[2]{Prop.~\ref{prop:#1}.\ref{p:#1-#2}} 

\newcommand{\refsect}[1]{Sect.~\ref{sect:#1}}

\renewcommand{\refeq}[1]{(\ref{eq:#1})} 
\newcommand{\reffig}[1]{Fig.~\ref{fig:#1}}

\newcommand{\ie}{\textit{i.e.}\xspace}

\newcommand{\ih}{\textit{i.h.}\xspace}

\newcommand{\leftsh}{\text{left}\xspace}
\newcommand{\rightsh}{\text{right}\xspace}

\newcommand{\red}[1]{{\color{red} {#1}}}
\newcommand{\blue}[1]{{\color{blue} {#1}}}

\newcommand{\purple}[1]{{\color{purple} {#1}}}

\newcommand{\defeq}{\coloneqq} 
\newcommand{\grameq}{\Coloneqq} 
\newcommand{\set}[1]{\{#1\}}

\newcommand{\card}[1]{\# #1}

\newcommand{\nat}{\mathbb{N}}
\newcommand{\size}[1]{|#1|}
\newcommand{\sizectx}[1]{\size{\typelist{#1}}}

\newcommand{\sizetyp}[1]{\size{#1}}

\renewcommand{\l}{\lambda}
\newcommand{\isub}[2]{\{#1/#2\}}
\newcommand{\replace}[2]{#1{\shortleftarrow}#2}
\renewcommand{\isub}[2]{\{\replace{#1}{#2}\}}

\newcommand{\fv}[1]{{\sf fv}(#1)}

\newcommand{\Rew}[1]{\rightarrow_{#1}}

\newcommand{\tob}{\Rew{\beta}}

\newcommand{\shufeqext}{\shufeqext} 

\newcommand{\tm}{t}
\newcommand{\tmtwo}{u}
\newcommand{\tmthree}{r}

\newcommand{\tmp}{\tm'}

\newcommand{\var}{x}
\newcommand{\vartwo}{y}
\newcommand{\varthree}{z}

\newcommand{\arbctxp}[1]{\arbctxp{#1}}
\newcommand{\arbctxtwop}[1]{\arbctxtwop{#1}}

\ifthenelse{\boolean{acmmart}
}{
  
  }{
  
  }

\newcommand{\deriv}{d}

\ifthenelse{\boolean{IEEEstyle}}{
    \newtheorem{theorem}{Theorem}[section]
    \newtheorem{lemma}[theorem]{Lemma}

    \newtheorem{proposition}[theorem]{Proposition}
    \newtheorem{definition}[theorem]{Definition}
}{}

\newcommand{\la}[1]{\lambda #1.}

\newcommand{\myproof}[1]{
\ifthenelse{\boolean{omitproofs}}{\begin{IEEEproof} Proof available but omitted for readability. \end{IEEEproof}}{#1}}

\newcommand{\tolo}{\Rew{\tiny\mathtt{lo}}}

\newcommand{\withproofs}[1]{\ifthenelse{\boolean{withproofs}}{#1}{}}

\newcommand{\withoutproofs}[1]{\ifthenelse{\boolean{withproofs}}{}{#1}}

\newcommand{\NoteProof}[1]{
	\marginnote{{\normalfont\scriptsize{Proof\,p.\,{\pageref{#1}}\,}}}}
\newcommand{\NoteState}[1]{
	\marginnote{{\normalfont\scriptsize{See p.\,{\pageref{#1}}}}}}
\renewcommand{\NoteProof}[1]{\marginnote{{Proof\,p.\,{\pageref{#1}}}}}
\renewcommand{\NoteState}[1]{\marginnote{{See\,p.\,{\pageref{#1}}\\\cref{#1}}}}

\crefname{proposition}{Prop.}{Props.}
\crefname{theorem}{Thm.}{Thms.}
\crefname{lemma}{Lemma}{Lemmas}
\crefname{corollary}{Cor.}{Cors.}
\crefname{section}{Sect.}{Sects.}
\Crefname{section}{Section}{Sections}

\newcommand{\doubt}[1]{}

\newcounter{numberone}
\newcounter{numberoneroman}
\newcounter{numberonealph}

\newcommand{\expr}{e}
\newcommand{\mset}[1]{\blue{[}#1\blue{]}}
\newcommand{\emptymset}{\mset{\,}}
\renewcommand{\emptymset}{\zero}
\newcommand{\zero}{\blue{\mathbf{0}}}

\newcommand{\ltype}{\red{\typefont{L}}}
\newcommand{\ltypetwo}{\ltype\red{'}}
\newcommand{\ltypethree}{\ltype\red{''}}

\newcommand{\typctx}{\blue{\Gamma}}
\newcommand{\typctxtwo}{\blue{\Delta}}
\newcommand{\typctxthree}{\blue{\Sigma}}

\newcommand{\tder}{\pi}

\newcommand{\hastype}{\!:\!}

\newcommand{\domain}[1]{\mathsf{dom}(#1)}
\newcommand{\typelist}[1]{\hat{#1}}

\newcommand{\mytr}[1]{\underline{#1}}
\newcommand{\auxtr}[1]{\overline{#1}}

\makeatletter
\newcommand\Copy[2]{
        \marginpar{\scriptsize \ \ \hyperlink{hl-appendix-#1}{Proof p.\,{\pageref*{appendix-#1}}}}
	\immediate\write\@auxout{\unexpanded{\global\long\@namedef{mytext@#1}{#2}
  }}%
	#2%
}

\newcommand\Paste[1]{%
        \hypertarget{hl-appendix-#1}{}\label{appendix-#1}
	\renewcommand{\inappendix}[1]{}
	\ifcsname mytext@#1\endcsname
	\@nameuse{mytext@#1}%
	\else
	``??''
	\fi
	\renewcommand{\inappendix}[1]{#1}
}
\makeatother

\newcommand{\inappendix}[1]{#1}

\newcommand{\disj}{\mathrel{\#}}

\newcommand{\larrow}[2]{\tarrow{#1}{#2}}

\newcommand{\derives}{\vartriangleright}
\newcommand{\derive}[2]{#1 \derives #2}
\newcommand{\concl}[4]{\derive{#1}{#2 \vdash #3 \hastype #4}}

\newcommand\Crumb\mytr
\newcommand\CrumbAux\auxtr

\makeatletter
\newcommand{\exder}{%
  \def\exderW[##1]{\triangleright_{##1}\ }%
  \def\exderWO{\triangleright\ }%
  \@ifnextchar[\exderW\exderWO%
  }
\makeatother

\newcommand{\tderiv}{\Phi}
\newcommand{\tderivtwo}{\Psi}
\newcommand{\tderivthree}{\Theta}

\newcommand{\typefont}[1]{{\mathsf{#1}}}
\newcommand{\mtype}{\blue{\typefont{M}}}
\newcommand{\mtypetwo}{\blue{\typefont{N}}}

\newcommand{\type}{\purple{\typefont{T}}}
\newcommand{\typetwo}{\type'}

\newcommand\mplus{\uplus}

\newcommand{\tyjp}[4]{{#3} \vdash^{#1} #2 \hastype #4}
\newcommand{\namedtyjp}[5]{#1 \vartriangleright \tyjp{#2}{#3}{#4}{#5}}

\newcommand{\dom}[1]{\mathsf{dom}(#1)}

\newcommand{\ruleAx}{\mathsf{ax}}
\newcommand{\ruleApp}{@}
\newcommand{\ruleAp}{\ruleApp}
\newcommand{\ruleFun}{\lambda}
\newcommand{\ruleMany}{\mathsf{many}}

\newcommand{\ruleAxPr}{\ruleAx^*}
\newcommand{\ruleApPr}{\ruleAp^*}
\newcommand{\ruleFunPr}{\ruleFun^*}
\newcommand{\ruleManyPr}{\ruleMany^*}

\newcommand{\ruleManyVar}{\ruleMany}
\newcommand{\ruleManyVal}{\ruleMany}

\newcommand{\I}{I}

\newcommand{\iI}{{i \in \I}}

\newcommand{\tarrow}[2]{#1 \,\red{\multimap}\, #2}

\newcommand{\mult}[1]{\mset{#1}}

\newcommand{\bigmplus}{\biguplus}

\newcommand{\Id}{{\mathsf{I}}}

\newcommand{\unitarysym}{{\textsc{u}}}

\newcommand{\rightsym}{{\textsc{r}}}
\newcommand{\rltype}{\ltype^{\rightsym}}

\newcommand{\rmtype}{\mtype^{\rightsym}}

\newcommand{\leftsym}{{\textsc{l}}}
\newcommand{\lltype}{\ltype^\leftsym}

\newcommand{\lmtype}{\mtype^\leftsym}

\newcommand{\urltype}{\ltype^{\unitarysym\rightsym}}

\newcommand{\urmtype}{\mtype^{\unitarysym\rightsym}}

\newcommand{\ulltype}{\ltype^{\unitarysym\leftsym}}

\newcommand{\ulmtype}{\mtype^{\unitarysym\leftsym}}

\newcommand\mydots{\hbox to .6em{.\hss.}}

\newcommand{\tderiveq}{\sim}

\newcommand{\Terms}{\Lambda}

\newcommand{\neu}{n}

\newcommand{\nf}{f}
\newcommand{\nfsym}{\mathtt{nf}}
\newcommand{\nftwo}{\nf'}
\newcommand{\nfthree}{\nf''}

\newcommand{\hnf}{h}

\newcommand{\hcompairs}[2]{\mathtt{ComPairs}(#1,#2)}
\newcommand{\hsubcompairs}[2]{\mathtt{ComPairsSub}(#1,#2)}

\newcommand{\xcompairs}[2]{\mathtt{ShComPairs}(#1,#2)}
\newcommand{\xsubcompairs}[2]{\mathtt{ShComPairsSub}(#1,#2)}

\newcommand{\sm}{\setminus \!\!\! \setminus}

\newcommand{\tvars}{\mathtt{TyVars}}
\newcommand{\tvarsp}[1]{\tvars(#1)}
\newcommand{\tvarsfp}[1]{\mathtt{TyVarsF}(#1)}
\newcommand{\tvar}{\red{\typefont{X}}}
\newcommand{\tvartwo}{\red{\typefont{Y}}}
\newcommand{\tvarthree}{\red{\typefont{Z}}}
\newcommand{\tvarfour}{\red{\typefont{W}}}

\newcommand{\vdashpr}{\vdash^{\mathtt{d}}}
\newcommand{\vdashun}{\vdash^1}

\newcommand\Deribbase[5]{{#3}\ {\pmb\vdash}_{#2}^{#1} {#4}\  {:}\  {#5}}

\makeatletter

\newcommand{\Deribase}[1]{%
  \def\DeribW[##1]{\Deribbase{##1}{#1}}%
  \def\DeribWO{\Deribbase{}{#1}}%
  \@ifnextchar[\DeribW\DeribWO%
  }

  \newcommand{\Deri}{%
  \def\DeriW_##1{\Deribase{##1}}%
  \def\DeriWO{\Deribase{}}%
  \@ifnextchar_\DeriW\DeriWO%
  }
\makeatother

\newcommand{\M}{\mtype}

\newcommand{\emptyctx}{\epsilon}

\newcommand{\col}{ : }

\newcommand{\sksize}[1]{\insize{#1}}
\newcommand{\hdsize}[1]{| #1|_{\httsym} }

\makeatletter

\newcommand{\Rewbase}{%
  \def\RewbaseW[##1]##2##3{\ {\xrightarrow{##1}}{}_{##2}^{##3}\ }%
  \def\RewbaseWO##1##2{\ {\xrightarrow{}}{}_{##1}^{##2}\ }%
  \@ifnextchar[\RewbaseW\RewbaseWO%
  }

\newcommand{\refequa}[1]{\refeq{#1}}

\newcommand{\httsym}{\mathtt{h}}
\newcommand{\toh}{\Rew{\httsym}}

\newcommand{\insize}[1]{\size{#1}_{\mathtt{in}}}

\newcommand{\compair}{p}
\newcommand{\compairtwo}{\compair'}

\definecolor{dgreen}{rgb}{0.0, 0.5, 0.0}

\newcommand{\perpRuleAx}{ax}
\newcommand{\perpRuleLambda}{\l}
\newcommand{\perpRuleAppR}{@r}
\newcommand{\perpRuleAppL}{@l}
\newcommand{\techreport}[1]{\ifthenelse{\boolean{techreport}}{#1}{}}
\newcommand{\paper}[1]{\ifthenelse{\boolean{techreport}}{}{#1}}

%% file: 00-Abstract.tex
\begin{abstract}
Intersection types are a standard tool in operational and semantical studies of the $\lambda$-calculus. De Carvalho showed how multi types, a quantitative variant of intersection types providing a handy presentation of the relational denotational model, allows one to extract precise bounds on the number of $\beta$-steps and the size of normal forms. 

In the last few years, de Carvalho's work has been extended and adapted to a number of $\lambda$-calculi, evaluation strategies, and abstract machines. These works, however, only adapt the first part of his work, that extracts bounds from multi type \emph{derivations}, while never consider the second part, which deals with extracting bounds from the multi types themselves. The reason is that this second part is more technical, and requires to reason up to \emph{type substitutions}. It is however also the most interesting, because it shows that the bounding power is \emph{inherent} to the relational model (which is induced by the types, without the derivations), independently of its presentation as a type system. 

Here we dissect and clarify the second part of de Carvalho's work, establishing a link with principal multi types, and isolating a key property independent of type substitutions.
\end{abstract}

%% file: 01-Introduction.tex
\section{Introduction}
\label{sect:intro}

Denotational semantics studies invariants of program evaluation. The typical way in which it is connected to the operational semantics of $\l$-calculi is at the \emph{qualitative} level, via \emph{adequacy}: the denotational interpretation $\sem\tm$ of a $\l$-term $\tm$ is non-trivial (typically non-empty) if and only if the evaluation~of~$\tm$~terminates.

At first sight, denotational semantics cannot provide \emph{quantitative} operational insights such as evaluation lengths, because of its invariance by evaluation. Things are in fact not so black and white. Being invariant by evaluation, denotational semantics models normal forms, 
and in a \emph{compositional} way: 
by composing the interpretations of two terms one can obtain the interpretation of the result of their application---therefore, denotational semantics does reflect the evaluation process~\emph{somehow}.

The aim of this paper is to revisit some overlooked---but we believe important---results for the $\l$-calculus by de Carvalho, about the extraction of bounds on evaluation lengths and the size of normal forms from the interpretation of terms into the relational model.

\subparagraph{Relational Semantics and Multi Types.}
The relational model \cite{Girard88,DBLP:journals/apal/BucciarelliE01} is a simple denotational semantics of the $\l$-calculus induced by the relational model of linear logic, via the representation of the  $\l$-calculus in linear logic. It is a paradigmatic model, underlying many others \cite{DBLP:journals/mscs/Ehrhard93,DBLP:journals/mscs/Ehrhard05,DBLP:journals/iandc/DanosE11,DBLP:conf/lics/LairdMMP13,DBLP:conf/lics/Ong17}, mainly studied by Ehrhard and his students and co-authors \cite{HYLAND200643,DBLP:conf/lfcs/BucciarelliEM09,DBLP:conf/csl/CarraroES10,DBLP:conf/csl/BucciarelliCEM11,DBLP:conf/csl/Ehrhard12,DBLP:journals/tcs/Ehrhard12,DBLP:conf/fossacs/Ehrhard20,DBLP:conf/csl/Carvalho16,DBLP:journals/entcs/ManzonettoR14,DBLP:conf/mfcs/Manzonetto09,DBLP:journals/mscs/PaoliniPR17,DBLP:journals/lmcs/BreuvartMR18}, the importance of which has emerged slowly.  One of its features is that it admits a handy presentation via an intersection type system.

The distinguished property of intersection types is that they \emph{characterize} termination properties, in the sense that they not only enforce termination, but they also type \emph{all} terminating terms. Additionally, by tuning the definition of the type system, one can capture different notions of termination (weak head/head/leftmost termination, strong normalization, call-by-value/need, etc). Such a characterization usually induces a model: the set of types for a term $\tm$ is an invariant of the characterized notion of evaluation $\to$, which gives rise to a denotational semantics which is adequate for $\to$. Therefore, intersection type systems are a tool halfway the operational and the denotational semantics of the $\l$-calculus.

Multi types are a \emph{linear logic-related} variant of intersection types where intersections are \emph{non-idempotent} (they are also known as \emph{non-idempotent intersection types}), that is where $A \cap A \neq A$. The set of multi types judgements derivable for a term $\tm$ provides a denotation $\sem\tm$ which coincides with the interpretation into the relational model. 

\subparagraph{Multi Types and Quantitative Bounds.}
In a seminal work, de Carvalho used the multi type system to obtain exact bounds on the evaluation lengths and the size of normal form for $\l$-terms \cite{Carvalho07,deCarvalho18}. The relevance of his work was fully appreciated by the community only a decade later (as surveyed below), when it blossomed into a number of variations and extensions by other authors. De Carvalho developed \emph{two} main results, but only the first one has widely permeated the community. The second one is arguably the most technical but also the deeper one. The aim of this paper is to make it accessible to a wider audience. 

De Carvalho's original presentation in \cite{Carvalho07,deCarvalho18} uses multi types to measure two forms of strong evaluations realized by abstract machines, implementing the head and leftmost call-by-name strategies. In this overview, we prefer to slightly depart from the (by now somewhat outdated) details of his work, forgetting about abstract machines, focussing on leftmost evaluation (the head case is treated at the end of the paper), and isolating three (rather than two) kinds of bounds:
\begin{enumerate}
\item \emph{Bounds from type derivations}. The size $\size\tderiv$ of a type derivation $\concl{\tderiv}{\typctx}{\tm}{\ltype}$ bounds the number of leftmost steps from $\tm$ to its normal form $\nfsym{(\tm)}$ \emph{plus} the size $\size{\nfsym{(\tm)}}$ of the normal form. Moreover, derivations of minimal size provide exact bounds. 

\item \emph{Size bounds from types}. The types in the final judgment---
a point of the relational interpretation $\sem\tm$---also provide a bound, independently of the derivation $\tderiv$. Being invariant by evaluation, they cannot bound evaluation lengths. They do however bound the size $\size{\nfsym{(\tm)}}$ of the normal form, and bounds are exact when types are minimal.

\item \emph{Bounds from composable types}. De Carvalho shows that types can be used to bound evaluation lengths \emph{compositionally}: from the types in $\sem\tm$ and $\sem\tmtwo$ it is possible to extract bounds about the leftmost evaluation and the normal form of $\tm\tmtwo$, with \emph{no reference to type derivations}.  Exact bounds rest on a complex construction involving type substitutions.
\end{enumerate}
The third kind of bounds is de Carvalho's second result, and it is where the bounding power of the multi type system (which is 
just one possible way of defining relational semantics) is \emph{lifted} to relational semantics. Therefore, the lifting guarantees that the bounding power is an \emph{inherent} feature of the relational model---multi types are just a handy tool to show it.

Of the three results, the third one is the most technical. In particular, it requires to enrich the type system with an infinity of type variables and work up to type substitutions. The puzzling fact is that  these extra features play no role in the  two previous points.

\subparagraph{De Carvalho's Legacy.} De Carvalho developed his results in his PhD, defended in 2007 \cite{Carvalho07}. His work was known by the community thanks to a technical report, turned into a journal paper only much later, in 2018 \cite{deCarvalho18}.
Soon after his PhD, he adapted his work to linear logic, 
with Pagani and Tortora de Falco \cite{DBLP:journals/tcs/CarvalhoPF11,DBLP:journals/iandc/CarvalhoF16}. 
Then, Bernadet and Graham-Lengrand adapted his work to measure the longest evaluation in the $\l$-calculus \cite{DBLP:journals/corr/BernadetL13}, but they only provided bounds of the first kind (\emph{from type~derivations}).

At the time, it was not known whether it would make sense to count the number of $\beta$-steps (or linear logic cut-elimination steps) as a reasonable measure of complexity. After the case of the $\l$-calculus was clarified in the positive by Accattoli and Dal Lago \cite{DBLP:journals/corr/AccattoliL16}, de Carvalho's work has been revisited by Accattoli et al. \cite{DBLP:journals/jfp/AccattoliGK20}. The revisitation started a new wave of works adapting de Carvalho's study to many evaluation strategies and extensions of the $\l$-calculus, including call-by-value \cite{DBLP:conf/aplas/AccattoliG18,DBLP:journals/pacmpl/AccattoliG22}, call-by-need \cite{DBLP:conf/esop/AccattoliGL19}, a linear logic presentation of call-by-push-value \cite{DBLP:conf/flops/BucciarelliKRV20,DBLP:conf/csl/KesnerV22}, the $\lambda\mu$-calculus \cite{DBLP:conf/lics/KesnerV20}, the $\l$-calculus with pattern matching \cite{DBLP:conf/types/AlvesKV19}, generalized applications \cite{DBLP:conf/fossacs/SantoKP22}, fully lazy sharing \cite{DBLP:conf/fossacs/KesnerPV21}, global state \cite{DBLP:conf/wollic/AlvesKR23}, the probabilistic $\l$-calculus \cite{DBLP:journals/pacmpl/LagoFR21}, the abstract machine underlying the geometry of interaction \cite{DBLP:journals/pacmpl/AccattoliLV21}, and even adapted as to measure \emph{space} \cite{DBLP:conf/lics/AccattoliLV21,DBLP:journals/pacmpl/AccattoliLV22}. All these works provide bounds of the first kind, and some of them also of the second kind, but \emph{none of them} deals with those of the third kind (\emph{bounds from composable types}).

\subparagraph{Contributions.} This paper revisits the \emph{bounds from composable types}, 
appeared only in \cite{deCarvalho18,DBLP:journals/tcs/CarvalhoPF11}. Beyond providing cleaner proofs of the results, we have a very close look at these bounds, isolating the subtleties and decomposing 
the proofs in smaller steps. In particular: 
\begin{itemize}
	\item \emph{Subtlety 1, minimality does not work}: when bounding a single term, both derivations and types provide exact bounds when they are minimal. When dealing with the application of $\tm$ to $\tmtwo$, every pair of composable types for them provides bounds. We stress that the technicalities for bounds from composable types are inherent to the problem, since the minimal composable pair in general does \emph{not} provide exact bounds.
	\item \emph{Subtlety 2, the gap between derivations and types}: we draw attention to the fact that the previous subtlety stems from a fact about derivations in \emph{isolation}, namely that, for a normal term $\tm$, both the derivation $\concl{\tderiv}{\typctx}{\tm}{\ltype}$ and the types in $\typctx$ and $\ltype$ provide bounds for $\size\tm$, but they may not coincide. In general, the bound from types is \emph{laxer}. The bounds gap hinders the possibility of lifting the bounding power from derivations to types, if bounds from some derivations are not reflected by any type in the interpretation $\sem\tm$. 
	\item \emph{Dry representation and type substitutions}: we isolate the property behind de Carvalho's solution of this problem, which rests on type variables and type substitutions, and we connect it with the study of principal types. The idea is that given a type derivation $\concl{\tderiv}{\typctx}{\tm}{\ltype}$ for which there is a gap between the bound from $\tderiv$ and the bound from $\typctx$ and $\ltype$, there always is a second \emph{dry} derivation $\concl{\tderivtwo}{\typctxtwo}{\tm}{\ltypetwo}$, whose types $\typctxtwo$ and $\ltypetwo$ give the same bound as the first derivation $\tderiv$, \emph{plus} a type substitution $\sigma$ turning $\tderivtwo$ into $\tderiv$. Then, all bounds coming from derivations can also be seen as coming from types, potentially from the types of other derivations---the dry ones---but related via type substitutions.

	\item \emph{Removing substitutions}: lastly, we show that \emph{weaker} bounds from composable types can be obtained \emph{without} dealing with an infinity of type variables and type substitutions. This point provides both a simplified route to the (slightly weaker) bounds and an explanation of why these additional technicalities are needed for the full result.
\end{itemize}

\subparagraph{Internal \emph{vs} External View.} The problem studied in the paper can also be seen in a more abstract way. Terms, or more generally programs, can be studied from two perspectives, which are distinct and yet entangled. The internal view considers programs as closed systems and looks at their internal evaluation in isolation. In the external view, programs are seen as parts of larger systems. What is relevant is how programs compose and interact with each other, their internal evaluation is instead secondary and hidden. 

Cost analyses are usually done following the internal view, while denotational semantics and types are external-oriented tools. Normal forms can be seen as internalizing the external information, as they are normal for the internal process and thus retain only information for potential external interactions. Multi type derivations capture both the number of steps of typed terms (which is an internal information) and the structure of their normal forms (which is the internalization of external information). Multi types instead capture the external information only (each type capturing a potential interaction). 

Bounds from composable types connect internal and external information, as they use types (that are external) to bound the length of evaluation (which is internal) of the isolated system given by the two composed terms. The bounds are obtained building over the connection between types and normal forms. The difficulty in this study stems from the fact that in general there is a gap between how the external information is represented in types (richly, distinguishing between different interactions) and how it is internalized in normal forms (in a raw way, all possible interactions are flattened on a single object).

\subparagraph{Limits of de Carvalho's Approach.} The aim of this paper is also to highlight a fundamental limitation of de Carvalho's work. As we ourselves suggested, bounds from composable types can be seen as lifting the bounding power from the type system to the relational model. There is however a delicate point, as the lifting does not cover the whole of the model. The relational model, indeed, does not only interpret full normal forms and terminating terms, but also \emph{head} normal forms and terms that are \emph{head normalizable}. In particular, there are terms that are \emph{hereditarily} head normalizing and yet never fully normalize, such as fix-point combinators. For these terms, which have non-empty interpretation in the relational model, de Carvalho's study does not say anything, because it assumes that the terms to be composed are fully normal. De Carvalho's bounds for head reduction do not solve the issue, they rather make it worse, since his theorems for the head case still have to assume that the composed terms are fully normal, which seems an unnatural assumption that is nonetheless mandatory.

Consequently, de Carvalho's technique does not apply to all terms having non-empty interpretation in the relational model. Here, we point out the problem, which is a first essential step. We do not aim at solving it, because it seems to require the development of a new approach, not just a refinement of de Carvalho's. Abstractly, the limits of his technique come from the fact that external information is measured by reducing it to internal information, rather than measuring it \emph{separately} from it, which would allow one to measure external information even when the internal evaluation process does not fully terminate.

\subparagraph{Proofs.} A few proofs are omitted and can be found in the Appendix. 

%% file: 02-Reductions.tex
\section{Head and Leftmost Reductions and Normal Forms}
\label{sect:reductions}

\subparagraph{Basics of $\l$.}The set $\Terms$ of untyped $\lambda$-terms is defined by
$\tm \grameq \var \mid \la{\var}{\tm} \mid \tm\,\tm$, which are considered as quotiented by $\alpha$-equivalence.
The capture-avoiding substitution of $\var$ by $\tmtwo$ in $\tm$
is written $\tm\isub{\var}{\tmtwo}$.


\subparagraph{$\beta$-Reduction and Normal Forms.}  $\beta$-reduction ${\tob} \subseteq \Terms \times \Terms$ is defined as follows:
\begin{center}$
\begin{array}{c@{\hspace{1.1cm}}c@{\hspace{1.1cm}}c@{\hspace{1.1cm}}c}
\multicolumn{4}{c}{\textsc{$\beta$-Reduction}}
\\[5pt]
	\begin{prooftree}[separation=1em, label separation = .2em]
		\hypo{ }
		\infer1[\scriptsize$\perpRuleAx$]
		{(\la\var\tm) \tmtwo \tob \tm \isub\var \tmtwo}
	\end{prooftree}
 	&
	\begin{prooftree}[separation=1em, label separation = .2em]
		\hypo{\tm \tob \tm' }
		\infer1[\scriptsize$\perpRuleLambda$]{\la\var\tm\ \tob \la\var\tm' }
	\end{prooftree}
	&
	\begin{prooftree}[separation=1em, label separation = .2em]
		\hypo{\tm \tob \tm' }
		\infer1[\scriptsize$\perpRuleAppL$]
		{\tm  \tmtwo \tob \tm'  \tmtwo}
	\end{prooftree}
&
	\begin{prooftree}[separation=1em, label separation = .2em]
		\hypo{\tm \tob \tm' }
		\infer1[\scriptsize$\perpRuleAppR$]
		{ \tmtwo  \tm \tob \tmtwo  \tm' }
	\end{prooftree}
\end{array}
$\end{center}
It is well known that $\beta$-reduction is non-deterministic but confluent, that is, a term can have at most one normal form. Normal forms (for $\beta$) are described by the following grammar relying on the mutually inductive notion of neutral term:
\begin{center}
$\begin{array}{r@{\hspace{.5cm}} rcl@{\hspace{.5cm}}@{\hspace{.5cm}}@{\hspace{.5cm}} r@{\hspace{.5cm}} rcl}
\textsc{Neutral terms} & \neu & \grameq & \var \mid \neu \nf
&
\textsc{Normal forms} & \nf & \grameq &\neu \mid \la\var\nf
\end{array}$
\end{center}
An alternative streamlined definition of normal forms is $\nf\defeq \la{\var_1}\ldots\la{\var_n}(\vartwo \nf_1\ldots \nf_k)$ with $n,k\geq0$, $\vartwo$ possibly equal to one of the $\var_i$, and the terms $\nf_j$ normal forms themselves.

For our quantitative study, we need a notion of size of normal forms. We use the following \emph{inner} one, which counts the number of inner nodes of a term, when seen as a tree having variables as leaves, as it is the one that best matches what shall be measured via multi types.
\begin{definition}[Inner size]
The inner size of $\l$-terms is defined as follows:
\begin{center}$
\begin{array}{ccc@{\hspace{1.5cm}}ccc@{\hspace{1.5cm}}ccc}
 \sksize\var &\defeq& 0
&
\sksize{\la\var\tmtwo} &\defeq& \sksize\tmtwo + 1 
&
\sksize{\tmtwo \tmthree} &\defeq& \sksize\tmtwo+\sksize\tmthree + 1
\end{array}$
\end{center}
\end{definition}

\subparagraph{Head Reduction.} A deterministic notion of evaluation for $\l$-terms is \emph{head reduction}, which reduces only the left branch of a term, when seen as a tree. The definition follows (it is obtained by omitting rule $\perpRuleAppR$ in the definition of $\beta$, and constraining $\tm$ not to be an abstraction in rule $\perpRuleAppL$):
\begin{center}
$\begin{array}{c@{\hspace{1.5cm}}c@{\hspace{1.5cm}}c}
\multicolumn{3}{c}{\textsc{Head reduction}}
\\[5pt]
	\begin{prooftree}[separation=1em, label separation = .2em]
		\hypo{ }
		\infer1[\scriptsize$\perpRuleAx$]
		{(\la\var\tm) \tmtwo \toh \tm \isub\var \tmtwo}
	\end{prooftree}
 	&
	\begin{prooftree}[separation=1em, label separation = .2em]
		\hypo{\tm \toh \tm' }
		\infer1[\scriptsize$\perpRuleLambda$]{\la\var\tm\ \toh \la\var\tm' }
	\end{prooftree}
	&
	\begin{prooftree}[separation=1em, label separation = .2em]
		\hypo{\tm \toh \tm' }
		\hypo{\tm \neq \la\var\tmthree}
		\infer2[\scriptsize$\perpRuleAppL$]
		{\tm  \tmtwo \toh \tm'  \tmtwo}
	\end{prooftree}
\end{array}$
\end{center}
Let $\Id\defeq \la\varthree\varthree$ be the identity combinator, $\delta\defeq \la\var\var\var$ be the duplicator, and consider the following examples: $\delta\Id \tm \toh \Id\Id\tm$ and $\la\vartwo(\delta\Id) \toh \la\vartwo\Id\Id$ but $(\la\vartwo(\delta\Id))\tm \not\toh (\la\vartwo\Id\Id)\tm$ , as it rather reduces to $(\delta\Id)\isub\vartwo\tm = \delta\Id$. 

Head reduction might not compute normal forms, since it does not evaluate arguments. Its notion of normal form follows:
\begin{center}
$\begin{array}{llllll}
\textsc{Head normal forms} & \hnf & \grameq & \la{\var_1}\ldots\la{\var_n}(\vartwo \tm_1\ldots \tm_k) 
\\&&& \mbox{with }n,k\geq0\mbox{ and }\vartwo\mbox{ possibly equal to one of the $\var_i$.}
\end{array}$
\end{center}
We shall also need a notion of \emph{head size} for head normal forms defined as $\hdsize\hnf\defeq n+k$ if $\hnf = \la{\var_1}\ldots\la{\var_n}(\vartwo \tm_1\ldots \tm_k)$.

\subparagraph{Leftmost Reduction.} Leftmost-outermost reduction $\tolo$ (shortened to \emph{leftmost}) is a deterministic extension of head reduction as to reduce arguments and reach normal forms. The definition relies on the notion of neutral term $\neu$ used to describe normal forms.
\begin{center}$
\begin{array}{c@{\hspace{1.5cm}}c}
\multicolumn{2}{c}{\textsc{Leftmost(-Outermost) Reduction}}
	\\[5pt]
	\begin{prooftree}[separation=1em, label separation = .2em]
		\hypo{ }
		\infer1[\scriptsize$\perpRuleAx$]
		{(\la\var\tm) \tmtwo \tolo \tm \isub\var \tmtwo}
	\end{prooftree}
 	&
	\begin{prooftree}[separation=1em, label separation = .2em]
		\hypo{\tm \tolo \tm' }
		\infer1[\scriptsize$\perpRuleLambda$]{\la\var\tm\ \tolo \la\var\tm' }
	\end{prooftree}
	\\[13pt]
	\begin{prooftree}[separation=1em, label separation = .2em]
		\hypo{\tm \tolo \tm' }
		\hypo{\tm \neq \la\var\tmthree}
		\infer2[\scriptsize$\perpRuleAppL$]
		{\tm  \tmtwo \tolo \tm'  \tmtwo}
	\end{prooftree}
&
	\begin{prooftree}[separation=1em, label separation = .2em]
		\hypo{\neu\mbox{ is neutral}}
		\hypo{\tm \tolo \tm' }
		\infer2[\scriptsize$\perpRuleAppR$]
		{ \neu  \tm \tolo \neu  \tm' }
	\end{prooftree}
\end{array}
$\end{center}

Examples: $\var (\Id\Id) (\Id\Id) \tolo \var \Id (\Id\Id)$ but $\var (\Id\Id) (\Id\Id) \not \tolo \var (\Id\Id) \Id$ and $\delta (\Id\Id) (\Id\Id) \not \tolo \delta \Id (\Id\Id)$.

Leftmost normal forms are simply normal forms and---crucially---leftmost reduction is \emph{normalizing}, that is, if $\tm$ has a $\beta$-reduction $\tm \tob^*\nf$ to normal form then leftmost reduction reaches that normal form, that is, $\tm\tolo^*\nf$. For a recent simple proof of this classic result see Accattoli et al. \cite{DBLP:conf/aplas/AccattoliFG19}. 

%% file: 03-Types.tex
\section{Multi Types, Head Reduction, and Bounds From Type Derivations}
\label{sect:types}
In this section, we give our presentation of  de Carvalho's system of multi types, and recall some results from the literature. In particular, we recall the characterization of head reduction, how multi types induce the relational model, and the bounds that can be extracted from type derivations for the length of head evaluations and the head size of head normal forms.

\subparagraph{Multi Types.} There are two layers of types, \emph{linear} and \emph{multi types}, built over a countably infinite set of (linear) type variables: 
\begin{center}
$\begin{array}{r@{\hspace{.5cm}} rll}
\textsc{Linear type variables} & \tvars &\defeq& \set{\tvar,\tvartwo,\tvarthree, \tvarfour, \tvar_1, \tvartwo', \tvarthree_2,\ldots}
\\
\textsc{Linear types} & \ltype, \ltypetwo &\grameq &\tvar\in\tvars \mid \larrow{\mtype}{\ltype}
\\
\textsc{Multi types} & \mtype, \mtypetwo &\grameq &\mset{\ltype_1, \dots, \ltype_n} \quad n \in \nat
\\
\textsc{Generic types} & \type, \typetwo &\grameq &\ltype \mid \mtype
\end{array}$
\end{center}
where $\mset{\ltype_1, \dots, \ltype_n}$ is our notation for finite 
multisets. 
The \emph{empty} multi type $\mset{\,}$ obtained by taking $n = 0$ is also denoted by $\emptymset$. Often, multi types are presented using a single type variable $\tvar$ instead of countably many. Most results are unaffected, but we shall see that for de Carvalho's semantic bounds we need countably many type variables.

%
A multi type $\mset{\ltype_1, \dots, \ltype_n}$ has to be intended as a conjunction $\ltype_1 \land \dots \land 
\ltype_n$ of linear types $\ltype_1, \dots, \ltype_n$, for a commutative, associative, non-idempotent conjunction 
$\land$ (morally a tensor $\otimes$), of neutral element $\emptymset$.
The intuition is that a linear type corresponds to a single use of a term $\tm$, and that $\tm$ is typed with a 
multiset 
$\mtype$ of $n$ linear types if it is going to be used (at most) $n$ times, that is, if $\tm$ is part of a larger term $\tmtwo$, then a copy 
of $\tm$ shall end up in evaluation position during the evaluation of $\tmtwo$.

\input{image-typing-rules}

\subparagraph{Typing Rules.} The derivation rules for the multi types system are in \Cref{fig:types}.
\emph{Judgments} have shape $\typctx \vdash \tm \hastype \mtype$ or $\typctx \vdash \tm \hastype \ltype$ where $\tm$ is a term, $\mtype$ is a multi type, $\ltype$ is a 
linear type, and $\typctx$ is a \emph{type context}, that is, a total function from variables to multi types 
such that  $\domain{\typctx} \defeq \{\var \mid \typctx(\var) \neq \emptymset\}$ is finite, usually represented as $\var_1 \hastype \mtype_1, \dots, \var_n \hastype \mtype_n$ (for some $n \in 
\nat$) if $\dom{\typctx} \subseteq \{\var_1, \dots, \var_n\}$ and $\typctx(\var_i) = \mtype_{i}$ for all $1 \leq i \leq 
n$. 

The abstraction rule $\ruleFun$ uses the notation $\typctx\sm\var$ for the type context defined as $\typctx$ on every variable but possibly $\var$, for which $(\typctx\sm\var)(\var)=\zero$. It is a compact way to express the rule in both the cases $\var\in\dom\typctx$ and $\var\notin\dom\typctx$. Note that the application rule $\ruleAp$ requires the argument to be typed with a multi type $\mtype$, which is necessarily introduced by rule $\ruleMany$, the hypotheses of which are a multi set of derivations, indexed by a possibly empty set $I$. When $I$ is empty, the rule has no premises and can type every term. For instance, $\vdash \Omega\hastype\zero$ is derivable, but no linear type can be assigned to $\Omega$. Essentially, $\zero$ is the type of erasable terms, and every term is erasable in the $\l$-calculus.

\subparagraph{Technicalities about Types.} The type context $\typctx$ is \emph{empty} if $\dom{\typctx} = \emptyset$.  
\emph{Multi-set sum} $\mplus$ is extended to type contexts point-wise,
\ie\  $(\typctx \mplus \typctxtwo)(\var) \defeq \typctx(\var) \mplus \typctxtwo(\var)$ for each variable $\var$.
This notion is extended to a finite family of type contexts as expected, 
in particular $\bigmplus_{i \in J\!} \typctx_i$ is the empty context  when $J = \emptyset$.
Given two type contexts $\typctx$ and $\typctxtwo$ such that $\dom{\typctx} \cap \dom{\typctxtwo} = \emptyset$, the 
type 
context $\typctx, \typctxtwo$ is defined by $(\typctx, \typctxtwo)(\var) \defeq \typctx(\var)$ if $\var \in 
\dom{\typctx}$, $(\typctx, \typctxtwo)(\var) \defeq \typctxtwo(\var)$ if $\var \in \dom{\typctxtwo}$, and $(\typctx, 
\typctxtwo)(\var) \defeq \emptymset$ otherwise.
Note that $\typctx, \var \hastype \emptymset = \typctx$, where we implicitly assume $\var \notin \dom{\typctx}$. 

\subparagraph{Type Derivations.} We write $\concl{\tderiv}{\typctx}{\tm}{\type}$ if $\tderiv$ is a (\emph{type}) \emph{derivation} (\ie a tree constructed using the rules in \Cref{fig:types}) with conclusion the multi judgment $\typctx \vdash \tm \hastype \type$.
In particular,  we write $\concl{\tderiv}{\,}{\tm}{\type}$ when $\typctx$ is empty.
We write $\derive{\tderiv}{\tm}$ if $\concl{\tderiv}{\typctx}{\tm}{\type}$ for some type context $\typctx$ and some type $\type$. 

We need a notion of size of type derivations, which shall be used to extract bounds for the number of evaluation steps and the size of normal forms.

\begin{definition}[Inner size of derivations]
	Let $\tderiv$ be a type derivation. 
	The \emph{inner size} $\insize{\tderiv}$ of $\tderiv$ is the number of  occurrences of rules $\ruleFun$ and $\ruleAp$ in $\tderiv$.
\label{def:two-sizes}
\end{definition}

\subparagraph{Subject Reduction and Expansion, and Relational Semantics.} The first properties of the type system that we recall are subject reduction and expansion, which hold for \emph{every} $\beta$-step.

\begin{proposition}
	Let $\tm \tob \tm'$.
	\label{prop:qual-subject} 
	\begin{enumerate}
		\item \label{p:qual-subject-reduction}
		\emph{Subject reduction}: 	if $\namedtyjp{\tderiv}{}{\tm}{\typctx}{\ltype}$ then there is a derivation $\namedtyjp{\tderiv'}{}{\tm'}{\typctx}{\ltype}$ such that $\insize{\tderiv'} \leq \insize{\tderiv} $.  Moreover, if $\tm\toh\tm'$ then $\insize{\tderiv'} = \insize{\tderiv} -2$.
		
		\item \label{p:qual-subject-expansion}
		\emph{Subject expansion}: if $\namedtyjp{\tderiv'}{}{\tm'}{\typctx}{\ltype}$ then there is a derivation $\namedtyjp{\tderiv}{}{\tm}{\typctx}{\ltype}$.
	\end{enumerate}
\end{proposition}

Together, subject reduction and expansion state that type judgements are \emph{invariants} of $\beta$-reduction. Such invariants actually induce a denotational model of the $\l$-calculus, its (call-by-name) \emph{relational semantics}.
\footnotetext{More precisely, such a model is 
	the restriction of the relational model for lineal logic to the image of  Girard's \cite{DBLP:journals/tcs/Girard87} call-by-name translation $(A \Rightarrow B)^\mathsf{n} = ! A^\mathsf{n} \multimap B^\mathsf{n}$ of the intuitionistic arrow into linear logic.}

Let $\tm$ be a term and $\var_1, \dots, \var_n$ ($n \geq 0$) be pairwise distinct variables. 
The list $\vec{\var} = (\var_1, \dots, \var_n)$ is \emph{suitable for} $\tm$ if $\fv{\tm} \subseteq \{\var_1, \dots, \var_n\}$.
If $\vec{\var} = (\var_1, \dots, \var_n)$ is suitable for $\tm$, the \emph{relational semantics} $\sem{\tm}_{\vec{\var}}$ \emph{of} $\tm$ \emph{for} $\vec{\var}$ is defined by:
\begin{center}$\begin{array}{lllllll}
	\sem{\tm}_{\vec{\var}} &\defeq &\{((\mtype_1,\dots, \mtype_n),\ltype) \mid 
	\exists 
	\, 
	\concl{\tderiv}{\var_1 \hastype \mtype_1, \dots, \var_n \hastype \mtype_n}{\tm}{\ltype} \}.
\end{array}$\end{center}
The following property is an immediate corollary of subject reduction and expansion.

\begin{proposition}[Invariance]
	Let $\vec{\var}  = (\var_1, \dots, \var_n)$ be suitable for two terms $\tm$ and $\tmtwo$.
	If $\tm \tob \tmtwo$ then $\sem{\tm}_{\vec{\var}} = \sem{\tmtwo}_{\vec{\var}}$.
\end{proposition}

\subparagraph{Characterizing Head Termination.} Note the quantitative aspect of subject reduction (\refpropp{qual-subject}{reduction}), stating that the derivation size \emph{cannot increase} after a reduction step, and that it \emph{decreases} with head steps (of 2 because removing a head $\beta$ redex removes a $\ruleFun$ and a $\ruleAp$ rule). It does not say that it decreases at \emph{every} step because steps occurring in sub-terms typed with rule $\ruleMany$ might not change the size. For instance, if $\var \tm \to \var\tm'$ and $\tm$ is typed using a empty $\ruleMany$ rule (\ie with 0 premises), which is a sub-derivation of size 0, then also $\tm'$ is typed using a empty $\ruleMany$ rule, of size 0. Therefore, not all typable terms terminate, as for instance $\var\Omega$ is typable as follows, for any linear type $\ltype$, but it has no normal form:
\begin{equation}
\begin{prooftree}[separation=1em]
			\hypo{}
			\infer1[\footnotesize$\ruleAx$]{\var\hastype \mset{\larrow{\mset\zero}\ltype} \vdash \var\hastype \larrow{\mset\zero}\ltype }
			\hypo{}
			\infer1[\footnotesize$\ruleManyVar$]{\vdash \Omega\hastype \zero}	
			\infer2[\footnotesize$\ruleAp$]{\var\hastype \mset{\larrow{\mset\zero}\ltype} \vdash \var\Omega\hastype \ltype}
\end{prooftree}
\label{eq:var-omega-tderiv}
\end{equation}

Since the size of type derivations decreases at every head step, it provides a termination measure for the head reduction of typable terms. Therefore, typable terms are head terminating---this is also called \emph{correctness} of the type system (with respect to head reduction). Conversely, every head normal forms is typable. Additionally, one can show that the size of type derivations bounds the head size of head normal forms, and that there exists derivations having exactly the head size of the normal form, as in the example \refeq{var-omega-tderiv} above.

\begin{proposition}[Typability of head normal forms]
\label{prop:typability-hnf}
Let $\hnf$ be a head normal form. 
\begin{enumerate}
\item \label{p:typability-hnf-forall}
\emph{Lax bounds for all pairs}: if $\tderiv \derives \typctx\vdash \hnf\hastype \ltype$ then $\hdsize\hnf\leq\insize\tderiv$;

\item \label{p:typability-hnf-exists}
\emph{Existence and exact bounds}: there exists a derivation $\tderiv \derives \typctx\vdash \hnf\hastype \ltype$ such that $\hdsize\hnf = \insize\tderiv$.
\end{enumerate}
\end{proposition}

Typability of all head normal forms (\refpropp{typability-hnf}{exists}) together with subject expansion (\refpropp{qual-subject}{expansion}) implies the \emph{completeness} of the type system: every head terminating term is typable.

\begin{theorem}[Typability characterizes head normalization]
\label{thm:head-characterization}
\hfill
\begin{enumerate}
\item \emph{Correctness}: if $\derive{\tderiv}{\tm}$ then there exists a head normalizing evaluation $\tm \toh^n \hnf$ with $\hnf$ normal and $2n +\hdsize\hnf \leq \insize{\tderiv}$.

\item \emph{Completeness}: if $\tm \toh^n \hnf$ is a head normalizing evaluation,
	then there exists a derivation $\derive{\tderiv}{\tm}$. In particular, there is a derivation $\tderiv$ for which $2n+\hdsize\hnf = \insize\tderiv$.
\end{enumerate}
\end{theorem}
The quantitative bounds involve $2n$ rather than $n$ because every $\beta$ redex is typed in a type derivation $\tderiv$ using two rules ($\ruleFun$ and $\ruleAp$). The type derivation captures only the head size of the normal form, because in general it ignores arguments, and so it cannot catch the inner size. For instance, for the head normal form $\var\Omega$ of head size $\hdsize{\var\Omega}=1$ (but inner size $\insize{\var\Omega} = 6$) the derivation \refeq{var-omega-tderiv} has inner size 1.

The head characterization theorem implies the following property of the semantics.
\begin{theorem}[Adequacy of relational semantics for head reduction]
	Let $\vec{\var}  = (\var_1, \dots, \var_n)$ be suitable for $\tm$. Then $\sem{\tm}_{\vec{\var}}$ is non-empty if and only if $\tm$ is $\toh$-normalizing.	\label{thm:adequacy}
\end{theorem}

Summing up, multi types naturally model head reduction. De Carvalho's bounds from composable types, however, rest on normal forms, which are reached by leftmost reduction, rather than on head normal forms and head reduction. Therefore, the next section recalls how multi types relate to leftmost reduction and normal forms.

%% file: image-typing-rules.tex
\begin{figure}[t!]
\centering
\begin{tabular}{c@{\hspace{.4cm}} c@{\hspace{.4cm}} c@{\hspace{.4cm}} c}
	$\begin{prooftree}[label separation = .2em]
					\infer0
					[\scriptsize$\ruleAx$]
					{\var \hastype \mset\ltype \vdash \var \hastype \ltype}
			\end{prooftree}$
						&
			$\begin{prooftree}[label separation = .2em]
					\hypo{\tyjp{}{\tm}{\typctx}{\ltype}}
					\infer1[\scriptsize$\ruleFun$]
					{\tyjp{}{\la{\var}{\tm}}{\typctx \sm \var}{\tarrow{\typctx(\var)}{\ltype}}}
			\end{prooftree}$
&
			
						$\begin{prooftree}[separation=1em, label separation = .2em]
					\hypo{\typctx \vdash \tm \hastype \tarrow{\mtype\!}{\!\ltype} }
					\hypo{\typctxtwo \vdash \tmtwo \hastype \mtype}
					\infer2[\scriptsize$\ruleAp$]
					{\typctx \uplus \typctxtwo \vdash \tm\tmtwo \hastype \ltype}
			\end{prooftree}$

&
$\begin{prooftree}[separation=1em, label separation = .2em]
				\hypo{\left[\tyjp{}{\tm}{\typctx_{i}}{\ltype_{i}}\right]_{\iI}}
				\infer1
				[\scriptsize$\ruleMany$]
				{\tyjp{}{\tm}{\biguplus_{\iI}\typctx_{i} }{\mult{\ltype_{i}}_{\iI}}}
				\end{prooftree}$
		\end{tabular}
	\caption{De Carvalho's multi type system.}
	\label{fig:types}
\end{figure}

%% file: 04-Bounds_From_Derivations.tex
\section{Bounds From Derivations Via (Unitary) Shrinking}
\label{sect:bounds-derivations}
In this section, we recall how to extend the results of the previous section to leftmost reduction $\tolo$ and full normal forms, via the so called \emph{shrinking} constraint. We follow the presentation of Accattoli et al. \cite{DBLP:journals/jfp/AccattoliGK20} (removing some of the aspects of their work that are not relevant here), but the definition of shrinking judgements is standard and not due to \cite{DBLP:journals/jfp/AccattoliGK20}, see for instance Krivine's book 
\cite{DBLP:books/daglib/0071545}, de Carvalho \cite{DBLP:books/daglib/0071545,deCarvalho18}, Kesner and Ventura \cite{DBLP:conf/ifipTCS/KesnerV14}, or Bucciarelli et al. \cite{DBLP:journals/igpl/BucciarelliKV17}.

\subparagraph{The Need for Shrinking.} Consider the derivation of end
  sequent
$\var\hastype\mset{\tarrow\zero\ltype}\vdash \var\Omega \hastype \ltype$ in \refeq{var-omega-tderiv}.
Since $\var\Omega$ is $\tolo$-diverging, this derivation has to be
excluded somehow. The problem here is that since $\var$ has an
erasing type---that is an arrow type with $\zero$ on the left---then the
diverging subterm $\Omega$ does not get typed. Excluding the use of
$\zero$ is too drastic, because the paradigmatic erasing term
$\la\vartwo\var$ is normal and can be typed only with
$\Deri{\var:\mset{\ltype}}{\la\vartwo\var}{\tarrow\zero\ltype}$.

The idea is that only \emph{some} occurrences of $\zero$ are dangerous. The
given examples seem to suggest that if $\zero$ occurs on the right side
of $\vdash$ it is fine, while if it occurs in the typing context it is
not. Things are subtler. Extending example
\refequa{var-omega-tderiv} with an abstraction, one obtains
the $\tolo$-diverging term $\la\var\var\Omega$ and the typing
$\Deri{}{\la\var\var\Omega}{\tarrow{\mset{\tarrow\zero\ltype}}\ltype}$,
that show that $\zero$ can be dangerous also on the right of $\vdash$. The dangerous occurrences of $\zero$ turn out to be those on the left of an \emph{even number of arrows}, considering the $\vdash$ symbols as an arrow. This is formalized by the \emph{shrinking} constraint, which allows one to characterize leftmost termination.

\subparagraph{Defining Shrinking.} There are two mutually defined notions of shrinking types, \emph{left} and \emph{right}, the key point of which is that right multi types \emph{cannot be empty} (note $n \geq 1$), so that $\emptymset$ is forbidden on the left of top arrows $\multimap$ for left linear types. Their definition follows:
\begin{center}
		$\begin{array}{r@{\hspace{.5cm}} rcl}
				\multicolumn{4}{c}{\textsc{Left and right (shrinking) types}}
		\\[5pt]

		\text{Right  linear type } & \rltype &\grameq & \tvar\in\tvars \mid  \larrow{\lmtype}{\rltype}
		\\
		\text{Left  linear type } & \lltype &\grameq & \tvar\in\tvars \mid  \larrow{\rmtype}{\lltype}
		\\[5pt]
		\text{Right  multi type } & \rmtype &\grameq &	\mset{\rltype_1, \dots, \rltype_n} \ \ n \geq 1
		\\
		\text{Left  multi type } & \lmtype &\grameq &	\mset{\lltype_1, \dots, \lltype_n} \ \ n \geq 0
		\end{array}$
\end{center}
The notions extend to type contexts and to derivations as follows:
	\begin{itemize}
	\item A type context $\var_1 \hastype \mtype_1, \dots, \var_n \hastype \mtype_n$ is \emph{\leftsh}  if each $\mtype_i$ is \leftsh;
	\item A derivation $\concl{\tderiv}{\typctx}{\tm}{\ltype}$ is \emph{shrinking} if $\typctx$ is \leftsh and $\ltype$ is \rightsh.
\end{itemize}
For instance, $\mset{\tvar}$ is both \leftsh and \rightsh, while $\zero$ is \leftsh but not \rightsh, and $\mset{\larrow{\emptymset}{\tvar}}$ is \rightsh but not \leftsh. Note that the derivation in \refeq{var-omega-tderiv} is not shrinking. By adding the shrinking constraint, we can now characterize leftmost normalization with multi types, with quantitative bounds involving the inner size of the normal form.

\begin{theorem}[Shrinking typability characterizes leftmost normalization, \cite{DBLP:journals/jfp/AccattoliGK20}]
\label{thm:shrinking-characterization}
\hfill
\begin{enumerate}
\item \emph{Correctness}: if $\derive{\tderiv}{\tm}$ is a shrinking derivation,
	then there exists a normalizing evaluation $\tm \tolo^n \nf$ with $\nf$ normal and $2n +\sksize\nf \leq \insize{\tderiv}$.

\item \emph{Completeness}: if $\tm \tolo^* \nf$ is a normalizing evaluation,
	then there exists a shrinking derivation $\derive{\tderiv}{\tm}$.
\end{enumerate}
\end{theorem}

\subparagraph{Unitary Shrinking.} Shrinking is enough to ensure termination, but not to capture the exact number of steps to normal form together with the exact size of the normal form. The point is somewhat dual to shrinkingness, as it concerns arguments that have to be typed, but that should not be typed \emph{too many times}. Consider the evaluation $\vartwo(\Id \varthree) \tolo \vartwo\varthree$ that involves 1 leftmost step and produces a normal form of inner size 1. The following shrinking derivation types the argument $\Id \varthree$ twice (the easy derivation of $\varthree:\mset\tvar\vdash \Id\varthree\hastype\tvar$ of inner size 2 is omitted), instead of once, and it has size 5, instead of the required 3 (obtained as 2*1+1):
\begin{equation}
\begin{prooftree}[separation=1em]
			\hypo{}
			\infer1[\footnotesize$\ruleAx$]{\vartwo:\mset{\tarrow{\mset{\tvar,\tvar}}\tvartwo} \vdash \vartwo\hastype \tarrow{\mset{\tvar,\tvar } } \tvartwo}
			
			\hypo{[\varthree:\mset\tvar\vdash \Id\varthree\hastype\tvar]_{i=1,2}}
			\infer1[\footnotesize$\ruleManyVar$]{\varthree:\mset{\tvar,\tvar} \vdash \Id\varthree\hastype \mset{\tvar,\tvar}}	
			\infer2[\footnotesize$\ruleAp$]{\vartwo:\mset{\tarrow{\mset{\tvar,\tvar}}\tvartwo},\varthree:\mset{\tvar,\tvar} \vdash \vartwo(\Id \varthree) \hastype \tvartwo}
\end{prooftree}
\label{eq:motivating-unitary}
\end{equation}
To obtain exact bounds, one needs \emph{unitary shrinking} types and derivations, that type arguments of normal forms only once, obtained by constraining some multi-sets---the right ones---to be singletons. The definition follows:
\begin{center}
		$\begin{array}{r@{\hspace{.5cm}} rcl}
		\multicolumn{4}{c}{\textsc{Unitary left and right (shrinking) types}}
		\\[5pt]
		\text{Unitary right linear types} & \urltype &\grameq & \tvar\in\tvars \mid  \larrow{\ulmtype}{\urltype}
		\\
		\text{Unitary left linear types} & \ulltype &\grameq & \tvar\in\tvars \mid  \larrow{\urmtype}{\ulltype}
		\\[5pt]
		\text{Unitary right multi types} & \urmtype &\grameq &	\mset{\urltype} 
		\\
		\text{Unitary left multi types} & \ulmtype &\grameq &	\mset{\ulltype_1, \dots, \ulltype_n} \ \ n \geq 0

		\end{array}$
\end{center}
The notions extend to type contexts and to derivations as expected:
	\begin{itemize}
	\item A type context $\var_1 \hastype \mtype_1, \dots, \var_n \hastype \mtype_n$ is \emph{unitary \leftsh}  if each $\mtype_i$ is unitary \leftsh;
	\item A derivation $\concl{\tderiv}{\typctx}{\tm}{\ltype}$ is \emph{unitary shrinking} if $\typctx$ is unitary \leftsh and $\ltype$ is unitary \rightsh.
\end{itemize}
For instance, the derivation in \refeq{motivating-unitary} is not unitary shrinking, because the multi type $\mset{\tarrow{\mset{\tvar,\tvar}}\tvartwo}$ of $\vartwo$ is not unitary left, since $\mset{\tvar,\tvar}$ is not unitary right. A derivable unitary shrinking typing for $\vartwo (\Id\varthree)$ is $\vartwo:\mset{\tarrow{\mset{\tvar}}\tvartwo},\varthree:\mset{\tvar} \vdash \vartwo(\Id \varthree) \hastype \tvartwo$, obtained via a derivation of inner size 3.

The following refinement of the shrinking characterization theorem (\refthm{shrinking-characterization}) holds.

\begin{theorem}[Unitary shrinking typability measures leftmost evaluation, \cite{DBLP:journals/jfp/AccattoliGK20}]
\label{thm:unitary-shrinking-characterization}
\hfill
\begin{enumerate}
\item \emph{Correctness}: if $\derive{\tderiv}{\tm}$ is a unitary shrinking derivation,
	then there exists a normalizing evaluation $\tm \tolo^n \nf$ with $\nf$ normal and $2n +\sksize\nf = \insize{\tderiv}$.

\item \emph{Completeness}: if $\tm \tolo^* \nf$ is a normalizing evaluation,
	then there exists a unitary shrinking derivation $\derive{\tderiv}{\tm}$.
\end{enumerate}
\end{theorem}

\subparagraph{Normal Forms.} The proof of the last theorem rests on two properties of normal forms that it is useful to state explicitly, for comparison with the study of the next sections.
\begin{proposition}[Unitary shrinking derivations and normal forms, \cite{DBLP:journals/jfp/AccattoliGK20}]
    \label{prop:shrinking-size-bound}
    Let $\nf$ be normal.
      \begin{enumerate}
  \item \label{p:shrinking-nfs-bounds-from-derivations-forall}
\emph{Lax bounds}: if $\tderiv \derives \nf$ is a shrinking derivation then $\sksize\nf\leq\insize\tderiv$;

\item \label{p:shrinking-nfs-bounds-from-derivations-exists}
\emph{Exact bounds}: there exists a unitary shrinking derivation $\tderiv \derives  \nf$ such that $\sksize\nf=\insize\tderiv$.
 \end{enumerate}
  \end{proposition}

%% file: 05-Bounds_From_Types.tex
\section{Bounds from Types} 
\label{sect:bounds-types}
In this section, we recall the bounds on the size of normal forms that can be extracted from \emph{types} rather than from \emph{type derivations}.

\subparagraph{Bounds from Types.} 
The types appearing in the final judgement of a shrinking derivation for $\tm$ bound the inner size $\sksize\nf$ of the normal
form $\nf$ of $\tm$, according to a notion of \emph{type size}
given below, and independently of the derivation itself. For example, consider the easily derivable (unitary shrinking) derivation $\tderiv \derives \vdash \delta\hastype \tarrow{\mult{\tarrow{\mult\tvar}\tvar, \tvar}}\tvar$ for $\delta=\la\var\var\var$.
There are two arrows in the type (judgement) and the normal form has inner size 
two. Of course, one also has to take into account the arrow symbols appearing in the 
typing context, when present.

Note, however, that types---even unitary shrinking ones---in general do not provide exact bounds: taking the 
derivation of $\tderiv$ for $\delta$ and substituting $\tvar$ with $\tarrow{\mult\tvartwo}\tvartwo$ everywhere in $\tderiv$ one obtains a unitary shrinking derivation $\tderivtwo$ having the same size of $\tderiv$ but final (still unitary shrinking) judgement:
\begin{center}$\tderivtwo \derives \Deri{}{\delta}{\tarrow{\mult{\tarrow{\mult{\tarrow{\mult\tvartwo}\tvartwo}}{
\tarrow{\mult\tvartwo}\tvartwo}, 
\tarrow{\mult\tvartwo}\tvartwo}}{\tarrow{\mult\tvartwo}\tvartwo}}$\end{center}
which has six arrows while $\insize{\delta}=2$.

\begin{definition}[Type size]
The  size $\size\cdot$ of types and  typing contexts is defined as follows:
\begin{center}$
\setlength{\arraycolsep}{3pt}
\begin{array}{r@{\hspace{.7cm}}rcl @{\hspace{.7cm}} rcl @{\hspace{.7cm}} rcl}
\textsc{Types} &
\size\tvar & \defeq & 0
&
\size{\tarrow\M\ltype} & \defeq & \size\M + \size\ltype + 1
&
\size{\mset{\ltype_1, \dots, \ltype_n}} & \defeq & \sum_{i=1}^n\size{\ltype_i}
\\
\textsc{Type ctxs} &
\size{\emptyctx} & \defeq & 0
&
\size{\var\col\M; \typctx} & \defeq & \size\M + \size\typctx
\end{array}$\end{center}
\end{definition}
Clearly, $\size{\type} \geq 0$ and $\size{\mtype} = 0$ if and only if $\mtype$ is a possibly empty multi set of type variables.

Given a type context $\typctx = \var_1 \hastype \mtype_1, \dots, \var_n \hastype \mtype_n$ we often consider the list of 
its types, 
noted  $\typelist\typctx \defeq (\mtype_1, \dots, \mtype_n)$.  Since any list of multi types $(\mtype_1, \dots, 
\mtype_n)$ can be seen as extracted from a type context $\typctx$, we 
use the notation $\typelist\typctx$ for lists of multi types.
The \emph{size} of a list of multi types is 
$\size{(\mtype_1, \dots, \mtype_n)} \allowbreak\defeq \allowbreak\sum_{i=1}^n \size{\mtype_i}$, and that of the conclusion 
of a derivation $\concl{\tder}{\typctx}{\expr}{\ltype}$ is $\size{(\typelist\typctx, \ltype)} \defeq 
\size{\typelist\typctx} + \size\ltype$.
Clearly, $\domain{\typctx} = \emptyset$ implies $\size{\typelist\typctx} = 0$.

\begin{proposition}[Shrinking types bound the size of normal forms, \cite{DBLP:journals/jfp/AccattoliGK20}]  
\label{prop:shrinking-nfs-bounds-from-types} 
Let $\nf$ be a normal form.
  \begin{enumerate}
  \item \label{p:shrinking-nfs-bounds-from-types-forall}
\emph{Lax bounds for all types}: if $\tderiv \derives \typctx\vdash \nf\hastype \ltype$ is a shrinking derivation then $\sksize\nf\leq\size{(\typelist\typctx,\ltype)}$;

\item \label{p:shrinking-nfs-bounds-from-types-exists}
\emph{Exact bounds for special types}: there exists a unitary shrinking derivation $\tderiv \derives \typctx\vdash \nf\hastype \ltype$ such that $\sksize\nf=\size{(\typelist\typctx,\ltype)}$.
 \end{enumerate}
\end{proposition}

%% file: 06-Dissecting_Bounds_From_Types.tex
\section{Dissecting Bounds From Types via Skeletons and Dry Judgements} 
\label{sect:dissecting}
In this section, we decompose and elaborate over the bounds on the size of normal forms extracted from types given in the previous section. The analysis is the main contribution of this paper. In particular, we develop notions and tools that shall be used in the next section to understand the issues concerning how to extract exact bounds from composable types.

\subparagraph{Types Bound the Size of Derivations.} The first observation is that the lax bounds of \refpropp{shrinking-nfs-bounds-from-types}{forall} are a consequence of the more general fact that types bound the size of derivations, proved next, together with the already proved fact that derivations bound the size of normal forms (\refprop{shrinking-size-bound}). The second point of the following proposition is the main statement, the first one is an auxiliary one that is needed for the proof to go through.
\begin{proposition}[Types bound the size of derivations for normal forms]
	\label{prop:types-bound-normal-derivations}
	Let $\concl{\tderiv}{\typctx}{\tm}{\type}$ be a derivation.
	\begin{enumerate}
		\item\label{pappendix:types-bound-normal-derivations-inert}\emph{Neutral:} if $\tm$ is a neutral term then $\insize\tderiv \leq \sizectx{\typctx} -\sizetyp{\type}$.
		\item\label{pappendix:types-bound-normal-derivations-fireball}\emph{Normal:} if $\tm$ is a normal form then 
$\insize\tderiv \leq \sizectx{\typctx} + \sizetyp{\type}$.
	\end{enumerate}
\end{proposition}

\input{\proofspath/strong/strong-type_size_forall}
Note that the bound holds for \emph{every} derivation, without requiring them to be shrinking. This fact means that the connection between types and derivations is stronger than the one between derivations and normal forms. Note also that the bound does \emph{not} hold for \emph{head} normal forms, as can be seen by inspecting examples \refeq{var-omega-tderiv} (p. \pageref{eq:var-omega-tderiv}) and \refeq{motivating-unitary} (p. \pageref{eq:motivating-unitary}).

\subparagraph*{Exact Bounds from Types.} We now turn our attention to exact bounds. Having showed that bounds for normal forms factor through bounds for derivations (\refprop{types-bound-normal-derivations}), we actually turn to exact bounds for \emph{derivations}, from types. The idea, as usual, is that exact bounds are given by types of minimal size. To describe such minimal types we shall use a modified \emph{dry} typing system for normal forms related to principal judgements, that shall derive only minimal types. Additionally, we use a relation between derivations having the same structure but assigning possibly different types, also considered by de Carvalho.

\subparagraph*{Skeleton Equivalence.} We formalize the notion of derivations having the same \emph{skeleton}, that is, the same \emph{mute} structure. Skeleton
equivalence $\tderiveq$
relates derivations having the same rules arranged in the same way, but not necessarily having the same types.

\begin{definition}[Skeleton equivalence]
Let $\tm$ be a term. 
Two derivations $\derive{\tderiv}{\tm}$ and $\derive{\tderivtwo}{\tm}$ 
are \emph{skeleton equivalent}, noted $\tderiv \tderiveq \tderivtwo$, if they end with the same kind of rule and the derivations on the 
premises are $\tderiveq$-equivalent, namely they fall in one of the following cases:
\begin{itemize}
	\item Both $\tderiv$ and $\tderivtwo$ are axioms.
	\item Both $\tderiv$ and $\tderivtwo$ end with rule $\ruleAp$, their two left premises $\tderiv_l$ and $\tderivtwo_l$ 
satisfy $\tderiv_l \tderiveq\tderivtwo_l$, and their right premises $\tderiv_r$ and $\tderivtwo_r$ satisfy $\tderiv_r 
\tderiveq\tderivtwo_r$---similarly for rules $\ruleFun$.
	\item Both $\tderiv$ and $\tderivtwo$ end with a rule $\ruleManyVal$ with $n$ premises and there is a permutation 
$\rho$ of $\set{1,\dots,n}$ such that the $i$-th premise $\tderiv_i$ of $\tderiv$ and the $\rho(i)$-th premise 
$\tderivtwo_{\rho(i)}$ of $\tderivtwo$ satisfy $\tderiv_i \tderiveq\tderivtwo_{\rho(i)}$ for 
$i\in\set{1,\dots,n}$. 
\end{itemize} 
\end{definition}

The next lemma shows that skeleton equivalence preserves more or less everything one can imagine, but types. We denote by $\card{m}$ the cardinality of a multiset~$m$.

\begin{lemma}[Skeletal invariants]
\label{l:skel-equiv-properties}
Let $\concl{\tderiv}{\typctx}{\tm}{\type}$ and $\concl{\tderivtwo}{\typctxtwo}{\tm}{\typetwo}$ be two derivations such 
that $\tderiv \tderiveq \tderivtwo$. Then $\insize\tderiv = \insize\tderivtwo$, $\dom\typctx = \dom\typctxtwo$, $\card{(\typctx(\var))} = \card{(\typctxtwo(\var))}$ 
for every variable $\var$, and $\type$ is a multi type if and only if $\typetwo$ is, and in that case $\card\type = \card\typetwo$. Moreover, $\tderiv$ is shrinking (resp. unitary shrinking) if and only if $\tderivtwo$ is.
\end{lemma}
\begin{proof}
Straightforward induction on $\tderiv$.
\end{proof}

\subparagraph{Principal and Dry Judgements.} Simple types admits \emph{principal judgements} (or typings), that is, for every term $\tm$ there exists a principal judgement $\Gamma \vdash \tm \hastype A$ such that for every other judgement $\Delta \vdash \tm \hastype B$ for $\tm$ there exists a type substitution $\sigma$ such that $\Gamma\sigma=\Delta$ and $A\sigma=B$. Multi types do not have principal judgements, since there is no \emph{single} judgement for a term that subsumes all others \emph{up to substitutions}. The literature has studied a weakened notion of principal judgement, subsuming all judgements up to substitution \emph{and} up to another (very technical) operation called \emph{expansion}, which---roughly---duplicates multi sets \cite{DBLP:journals/mlq/CoppoDV81,DBLP:journals/tcs/Rocca88,DBLP:journals/tcs/KfouryW04}.

What we are going to do next, intuitively, is following the other natural route when principal judgements do not exist: we study a notion of principal \emph{set} of special judgements for a term $\tm$, called \emph{dry} judgements, which are such that every ordinary judgement for $\tm$ can be seen as a dry judgement up to substitution. In fact, we only study this property for normal forms, and we also relate the derivations producing those judgements. We need some definitions.

\subparagraph*{Supports and Substitutions.} The \emph{support} of a type derivation $\concl{\tderiv}{\typctx}{\tm}{\type}$ is the set $\tvarsp\tderiv\defeq\set{\tvar\ |\ \tvar\mbox{ occurs in }\tderiv}$ of type variables appearing in $\tderiv$, and the \emph{final support} is the set $\tvarsfp\tderiv\defeq\set{\tvar\ |\ \tvar\mbox{ occurs in }\typctx\mbox{ or }\type}$ of type variables appearing in the last judgement of $\tderiv$. We write $\tderiv\disj \tderivtwo$ as a shortcut for $\tvarsp\tderiv \cap \tvarsp\tderivtwo = \emptyset$ and given $\set{\tderiv_i}_{\iI}$ we write $\disj_{\iI}\tderiv_i$ when $\tvarsp{\tderiv_h} \cap \tvarsp{\tderiv_k} = \emptyset$ for any two distinct $h,k\in I$.

 A type substitution $\sigma$ is a function from type variables to 
linear types that is the identity but a for finite number of type variables. It is extended to act on types, multi types, type contexts, and derivations as expected.

\input{image-principal-typing-rules}

\subparagraph{Dry Judgements.} Dry judgements for normal forms are derived using the rules in \reffig{dry-type-system}. There are three key points. Firstly, only normal forms are typable. Secondly, neutral terms are always typed with a type variable (which is minimal) and when they are applied (in rule $\ruleApPr$) their type is \emph{enlarged} on-the-fly via a type substitution $\isub\tvar{(\tarrow\mtype\tvartwo)}$ depending on the type of the argument. Thirdly, the system uses many type variables, and for the rules with more than one premise (\ie $\ruleApPr$ and $\ruleManyPr$) it requires them to have disjoint supports. This is where having countably many type variables plays a role, as having only a finite number would not allow one to prove the \emph{subsumption up to substitutions} property of dry derivations, given by \refthmp{dry_representation}{sub} below.

Dry derivations can be seen as standard derivations, as the second point of the next lemma states. It is obtained using the straightforward fact that standard derivations are stable by type substitutions. Note the use of skeleton equivalence.

\begin{lemma}
\label{l:dry_is_standard}
Let $\tm$ be a term and $\nf$ be a normal form.
\begin{enumerate}
\item \label{p:dry_is_standard-sub}
\emph{Substitutivity for standard}: if $\tderiv \derives \typctx \vdash \tm \hastype  \ltype$ then for any linear type $\ltypetwo$ there exists $\tderiv_{\isub\tvar\ltypetwo} \derives \typctx\isub\tvar\ltypetwo \vdash \tm \hastype \ltype\isub\tvar\ltypetwo$ such that $\tderiv \tderiveq \tderiv_{\isub\tvar\ltypetwo}$.

\item \label{p:dry_is_standard-proper}
\emph{Dry derivations are standard}: if $\tderiv \derives \typctx \vdashpr \nf \hastype \ltype$ then there exists $\tderivtwo \derives \typctx \vdash \nf \hastype \ltype$ such that $\tderiv \tderiveq \tderivtwo$.
\end{enumerate}
\end{lemma}

\input{\proofspath/dry_is_standard}

Next, we prove that types in dry judgements are always minimal and---crucially---capture the size of the derivation itself. This is obtained via a strong property for type variables in dry judgements, reminiscent of similar properties in multiplicative linear logic, and enforced by the requirements about disjoint supports in the derivation rules.

\begin{proposition}
\label{prop:dry_is_minimal}
Let $\nf$ be a normal form.
\begin{enumerate}
\item \label{p:dry_is_minimal-one}
\emph{Dry derivations and variable types occurrences}: if $\tderiv \derives \typctx \vdashpr \nf \hastype \type$ then $\tvar$ has exactly two occurrences in $(\typctx,\type)$ for every $\tvar\in\tvarsfp\tderiv$.

\item \emph{Dry derivations are minimal}: if $\tderiv \derives \typctx \vdashpr \nf \hastype \ltype$ then $\insize\tderiv = \size{(\typelist\typctx, \ltype)}$.
\end{enumerate}
\end{proposition}
\input{\proofspath/dry_is_minimal}

\subparagraph{Dry Representation.} We now prove the key property of the analysis, which also justifies seeing dry derivations as defining a set of principal judgements. The idea is that every (standard) derivation $\tderiv$ admits a dry derivation $\tderivtwo$ that is skeleton equivalent to $\tderiv$---which by the skeletal invariants above entails that they have the same size---and such that there is a substitution turning $\tderivtwo$ into $\tderiv$. We need an auxiliary lemma that shall also be useful in the next section, about renamings of dry derivations.
\begin{lemma}[Dry derivations are stable by renaming]
\label{l:dry-stabile-by-renaming}
Let $\tderiv \derives \typctx \vdashpr \nf \hastype \type$ be a dry type derivation $\tvarsp\tderiv=\set{\tvar_1,\ldots,\tvar_n}$ and $\tvartwo_1,\ldots,\tvartwo_n$ be distinct type variables $\tderiv$. Then the derivation $\tderiv_{\isub{\tvar_1,\ldots,\tvar_n}{\tvartwo_1,\ldots,\tvartwo_n}}$ obtained by simultaneously replacing $\tvar_i$ with $\tvartwo_i$ in $\tderiv$ for $i\in\set{1,\ldots,n}$ is a dry type derivation such that $\tderiv \tderiveq \tderiv_{\isub{\tvar_1,\ldots,\tvar_n}{\tvartwo_1,\ldots,\tvartwo_n}}$.
\end{lemma}
\begin{proof}
Straightforward induction on $\tderiv$.
\end{proof}

\begin{theorem}
\label{thm:dry_representation} 
Let $\nf$ be a normal form and $\tderiv \derives \typctx \vdash \nf \hastype \type$ be a type derivation.
\begin{enumerate}
\item \label{p:dry_representation-core}
\emph{Dry representation}: there exists a dry derivation $\tderivtwo \derives \typctxtwo \vdashpr \nf \hastype \typetwo$ such that $\tderiv \tderiveq \tderivtwo$; 
\item \label{p:dry_representation-sub}
\emph{Type substitution}: there exists a type substitution $\sigma$ such that $\typctxtwo\sigma=\typctx$ and $\typetwo\sigma=\type$.
\end{enumerate}
\end{theorem}

\input{\proofspath/dry_representation}

\subparagraph{Removing Substitutions.} In the previous theorem, the substitution part rests on the properties of dry derivations enabled by the countable number of type variables in the type system. We now show that a slightly weaker result is possible even with only one type variable and without dry derivations. The type substitution part shall not be recoverable, but the representation and the quantitative bounds are. 

Let $\typctx \vdashun \tm \hastype \ltype$ denote a (standard) type derivation built using only \emph{$1$-types}, that is, types built using a single fixed type variable $\tvar$. 
\begin{theorem}[Size representation]
\label{thm:size-representation}
\techreport{\NoteProof{thmappendix:size-representation}}
Let $\nf$ be a normal term and  $\tderiv \derives \typctx \vdash \nf \hastype \ltype$ be a derivation. Then there exists a derivation $\tderivtwo \derives \typctxtwo \vdashun \nf \hastype \ltypetwo$ such that $\tderivtwo \tderiveq \tderiv$ and $\insize\tderivtwo = \sizectx\typctxtwo + \size\ltypetwo$.
\end{theorem}
The proof of the theorem in fact requires a stronger statement for the induction to go through, in particular having a separate point about neutral terms, for which a stronger property holds. \techreport{It can be found in the Appendix}\paper{See the technical report \cite{accattoli2023semantic}}.

%

%% file: proofs/strong/strong-type_size_forall.tex
\begin{proof}
	By mutual induction on the definition of neutral and normal terms, followed by an induction on the type derivation $\tderiv$.
	\begin{enumerate}
	\item \emph{$\tm$ is a neutral term.} Cases of the last rule:
	\begin{itemize}
		\item \emph{Rule $\ruleMany$}. Then $\type$ is a multi type $\mtype= \mset{\ltype_i}_{\iI}$ and the last rule is necessarily $\ruleMany$. So, necessarily, for some finite set of indices $I$,
		\begin{center}
		$\tderiv = 
		\begin{prooftree}
		\hypo{\tderiv_i\derives \tyjp{}{\tm}{\typctx_i}{\ltype_i}}
		\delims{\left[}{\right]_{\iI}}
		\infer1[\footnotesize$\ruleMany$]{\tyjp{}{\tm}{\mplus_{\iI}\typctx_i}{\mset{\ltype_i}_{\iI}}}
		\end{prooftree}$
		\end{center}
		where $\typctx = \mplus_{\iI}\typctx_i$.
		By \ih (on $\tderiv_i$), $\insize{\tderiv_i} \leq \sizectx{\typctx_i} -\sizetyp{\ltype_i}$, thus $\insize{\tderiv}= \sum_{\iI}\insize{\tderiv_i} \leq \sum_{\iI}\sizectx{\typctx_i} -\sum_{\iI}\sizetyp{\ltype_i}= \sizectx{\mplus_{\iI}\typctx_i} -\sizetyp{\mset{\ltype_i}_{\iI}} = \sizectx{\typctx}- \sizetyp{\mtype}$.
		
		\item \emph{Rule $\ruleAx$}, that is, $\tm = \var$. Then:
		\begin{center}
		$\tderiv = 
		\begin{prooftree}
		\infer0[\footnotesize$\ruleAx$]{\tyjp{}{\var}{\var \hastype \mset{\ltype}}{\ltype}}
		\end{prooftree}$
		\end{center}
		where $\type = \ltype$ and $\typctx = \var \hastype \mset\ltype$.
		Since $\insize{\tderiv} = 0$ and $\sizetyp{\type}=\sizetyp{\ltype}=\sizetyp{\mset\ltype} = \sizectx{\typctx}$,
		then $\insize{\tderiv} = 0 = \sizectx{\typctx}- \sizetyp{\type}$.

		\item \emph{Rule $\ruleAp$}, that is, $\tm = \neu \nf$. 
		Then necessarily:
		\begin{center}
		$\tderiv = 
		\begin{prooftree}
		\hypo{\tderiv_{\neu} \derives \typctxtwo \vdash \neu \hastype \tarrow{\mtypetwo}{\ltype}}
		\hypo{\tderiv_{\nf} \derives \typctxthree \vdash \nf \hastype\mtypetwo}
		\infer2[\footnotesize$\ruleApp$]{\tyjp{}{\neu \nf}{\typctxtwo \mplus \typctxthree}{\ltype}}
		\end{prooftree}$
		\end{center}
		where $\type = \ltype$ and $\typctx = \typctxtwo \mplus \typctxthree$.
		By \ih (on the definition of neutral terms and normal forms), $\insize{\tderiv_{\neu}} \leq \sizectx{\typctxtwo} - \sizetyp{\larrow{\mtypetwo}{\ltype}} = 
\sizectx{\typctxtwo} - \sizetyp{\mtypetwo} - \sizetyp{\ltype} -1$  and $\insize{\tderiv_{\nf}} \leq 
\sizectx{\typctxthree} + \sizetyp{\mtypetwo}$. 
		Therefore, 
		\begin{center}$\begin{array}{rcllllllll}
		\insize{\tderiv} & = & \insize{\tderiv_{\neu}} + \insize{\tderiv_{\nf}} + 1 		
		& \leq_{\ih} &\sizectx{\typctxtwo} - \sizetyp{\mtypetwo} - \sizetyp{\ltype} -1+ \insize{\tderiv_{\nf}} + 1  
		\\
		& = &\sizectx{\typctxtwo} - \sizetyp{\mtypetwo} - \sizetyp{\ltype}+ \insize{\tderiv_{\nf}}   		
		& \leq_{\ih} & \sizectx{\typctxtwo} - \sizetyp{\mtypetwo} - \sizetyp{\ltype}+ \sizectx{\typctxthree} + 
\sizetyp{\mtypetwo}	
\\	
		& = & \sizectx{\typctxtwo} + \sizectx{\typctxthree}- \sizetyp{\ltype}
		& = & \sizectx{\typctx} - \sizetyp{\type}.
		\end{array}$\end{center}
\end{itemize}

\item \emph{$\tm$ is a normal form}. If $\tm$ is a neutral term, then the statement follows from Point 1, which stronger than the statement that we need to prove. Otherwise, $\tm$ is an abstraction. If the last rule is $\ruleMany$ then we reason exactly as for neutral terms. The remaining cases is when the last rule is $\ruleFun$, that is, $\tm = \la{\var}{\nf}$ and $\tderiv$ is necessarily of the form:
		\begin{center}
		$\begin{prooftree}
		\hypo{\tderiv'\derives \typctxtwo \vdash \nf \hastype \ltypetwo}
		
		\infer1[\footnotesize$\lambda$]{\typctxtwo \sm \var \vdash \la{\var}\nf \hastype \larrow{\typctxtwo(\var)}{\ltypetwo}}
\end{prooftree}$
		\end{center}
		where $\type = \larrow{\typctxtwo(\var)}{\ltypetwo}$ and $\typctx = \typctxtwo \sm \var$.
		By \ih, 
		\begin{center}$\begin{array}{rclllllllllll}
		\insize{\tderiv'}  \leq_{\ih} & \sizectx{\typctxtwo} + \sizetyp{\ltypetwo}
		& = & \sizectx{\typctxtwo\sm\var} + \sizetyp{\typctxtwo(\var)} + \sizetyp{\ltypetwo}
		& = & \sizectx{\typctxtwo\sm\var} + \sizetyp{\larrow{\typctxtwo(\var)}{\ltypetwo}}-1
		\end{array}$\end{center}
		Therefore, 
		\begin{center}$\begin{array}{rcllllll}
		\insize{\tderiv} & = & \insize{\tderiv'} + 1
		& \leq & \sizectx{\typctxtwo\sm\var} + \sizetyp{\larrow{\typctxtwo(\var)}{\ltypetwo}}-1 +1
		\\
		& = & \sizectx{\typctxtwo\sm\var} + \sizetyp{\larrow{\typctxtwo(\var)}{\ltypetwo}}
		& = & \sizectx{\typctx} + \sizetyp{\type}. & 
		\end{array}$\end{center}\qedhere
	\end{enumerate}
\end{proof}

%% file: image-principal-typing-rules.tex
\begin{figure}[t]
\centering
\begin{tabular}{c@{\hspace{.5cm}} c@{\hspace{.5cm}} c@{\hspace{.5cm}} c}
	$\begin{prooftree}[label separation = .2em]
					\hypo{}
					\infer1
					[\scriptsize$\ruleAxPr$]
					{\var \hastype \mset\tvar \vdashpr \var \hastype \tvar}
			\end{prooftree}$
						&

						$\begin{prooftree}[separation=1em, label separation = .2em]
					\hypo{\tderiv \derives \typctx \vdashpr \neu \hastype \tvar }
					\hypo{\tderivtwo \derives \typctxtwo \vdashpr \nf \hastype \mtype}
					\hypo{\tvartwo\mbox{ fresh, }\tderiv\disj\tderivtwo}
					\infer3[\scriptsize$\ruleApPr$]
					{(\typctx\isub\tvar{(\tarrow\mtype\tvartwo)} \uplus \typctxtwo) \vdashpr \neu \nf \hastype \tvartwo}
			\end{prooftree}$
			\\\\
						$\begin{prooftree}[label separation = .2em]
					\hypo{ \typctx \vdashpr \nf \hastype \ltype}
					\infer1[\scriptsize$\ruleFunPr$]
					{\typctx \sm \var \vdashpr \la\var\nf \hastype \tarrow{\typctx(\var)}\ltype}
			\end{prooftree}$
&
$\begin{prooftree}[separation=1em, label separation = .2em]
				\hypo{\left[\tderiv_i \exder \typctx_i \vdashpr \nf \hastype \ltype_i\right]_{\iI}}
				\hypo{\disj_{\iI}\tderiv_i}
				\infer2
				[\scriptsize$\ruleManyPr$]
				{ \biguplus_{\iI}\typctx_{i} \vdashpr \nf \hastype \mult{\ltype_{i}}_{\iI} }
				\end{prooftree}$

		\end{tabular}
	\caption{Dry multi type system for normal forms.}
	\label{fig:dry-type-system}
\end{figure}

%% file: proofs/dry_is_standard.tex
\begin{proof}
The first point is a straightforward induction on $\tderiv$. The second point is by induction on $\tderiv$. The only rule of the dry system that is not a rule of the standard system is $\ruleApPr$:
\begin{center}$\begin{prooftree}[separation=1em, label separation = .2em]
					\hypo{\tderiv_\neu \derives \typctx_\neu \vdashpr \neu \hastype \tvar }
					\hypo{\tderiv_{\nftwo} \derives \typctx_{\nftwo} \vdashpr \nftwo \hastype \mtype}
					\hypo{\tvartwo\mbox{ fresh, }\tderiv\disj\tderivtwo}
					\infer3[\scriptsize$\ruleApPr$]
					{\typctx_\neu\isub\tvar{(\tarrow\mtype\tvartwo)} \uplus \typctx_{\nftwo} \vdashpr \neu \nftwo \hastype \tvartwo}
			\end{prooftree}$\end{center}
			with $\nf = \neu\nftwo$, $\typctx = \typctx_\neu\isub\tvar{(\tarrow\mtype\tvartwo)} \uplus \typctx_{\nftwo}$, $\type=\tvartwo$.
By \ih, there exist $\tderivtwo_\neu \derives \typctx_\neu \vdash \neu \hastype \tvar$ and $\tderivtwo_{\nftwo} \derives \typctx_{\nftwo} \vdash \nftwo \hastype \mtype$. By Point 1, there exists a derivation: 
$\tderivtwo_\neu\isub\tvar{(\tarrow\mtype\tvartwo)} \derives \typctx_\neu\isub\tvar{(\tarrow\mtype\tvartwo)} \vdash \neu \hastype \tarrow\mtype\tvartwo$
Then we build $\tderivtwo$ as follows:
\begin{center}$\begin{prooftree}[separation=1em, label separation = .2em]
					\hypo{ \tderivtwo_\neu\isub\tvar{(\tarrow\mtype\tvartwo)} \derives \typctx_\neu\isub\tvar{(\tarrow\mtype\tvartwo)} \vdash \neu \hastype \tarrow\mtype\tvartwo }
					\hypo{ \tderivtwo_{\nftwo} \derives \typctx_{\nftwo} \vdash \nftwo \hastype \mtype }
					\infer2[\scriptsize$\ruleAp$]
					{\typctx_\neu\isub\tvar{(\tarrow\mtype\tvartwo)} \uplus \typctx_{\nftwo} \vdash \neu \nftwo \hastype \tvartwo}
			\end{prooftree}$\end{center}
	Skeleton equivalence of $\tderiv$ and $\tderivtwo$ follows immediately from the skeleton equivalences of the \ih plus the one of Point 1.
\end{proof}

%% file: proofs/dry_is_minimal.tex
\begin{proof}
\begin{enumerate}
\item By induction on $\tderiv$, looking at its last rule. For $\ruleAxPr$, the statement evidently holds, and for $\ruleFunPr$ and rule $\ruleManyPr$ it follows from the \ih, since these rules preserve  and do not introduce occurrences of type variable. If the last rule of $\tderiv$ is $\ruleApPr$, then $\tderiv$ has shape:
\begin{center}
$\tderiv =
\begin{prooftree}[separation=1em, label separation = .2em]
					\hypo{\tderiv_1 \derives \typctxtwo \vdashpr \neu \hastype \tvar }
					\hypo{\tderiv_2 \derives \typctxthree \vdashpr \nftwo \hastype \mtype}
					\hypo{\tvartwo\mbox{ fresh, }\tderiv\disj\tderivtwo}
					\infer3[\scriptsize$\ruleApPr$]
					{\typctxtwo\isub\tvar{(\tarrow\mtype\tvartwo)} \uplus \typctxthree \vdashpr \neu \nftwo \hastype \tvartwo}
			\end{prooftree}$
\end{center}
with $\nf = \neu\nftwo$, $\type = \tvartwo$,  and $\typctx = \typctxtwo\isub\tvar{(\tarrow\mtype\tvartwo)} \uplus \typctxthree$. 
By \ih, $\tvar$ occurs exactly once in $\typctxtwo$, thus $\tvartwo$ occurs exactly twice in $(\typctx,\type)$. Note that $\tvarsfp\tderiv = (\tvarsfp{\tderiv_1}\setminus\set\tvar)\cup\tvarsfp{\tderiv_2}\cup\set\tvartwo$. By \ih, each type variable in $\tvarsfp{\tderiv_1}\setminus\set\tvar$ (resp. $\tvarsfp{\tderiv_2}$) occurs exactly twice in $\typctxtwo$ (resp. $(\typctxthree, \mtype)$). Moreover, by hypothesis $\tvarsp{\tderiv_1}\cap\tvarsp{\tderiv_2} = \emptyset$, so each such type variable occurs exactly twice in $(\typctx,\type)$.

\item By induction on $\tderiv$, looking at its last rule. Cases:
\begin{itemize}
\item \emph{Rule $\ruleAxPr$}: the statement holds because $\insize\tderiv=0=\size{(\mset\tvar, \tvar)}$.
\item \emph{Rule $\ruleManyPr$}: it follows from the \ih
\item \emph{Rule $\ruleFunPr$}: it follows from the \ih because rule $\ruleFunPr$ add 1 to the size of the derivation, but the the size of the judgement also grows of 1, because of the introduced arrow.

\item \emph{Rule $\ruleApPr$}: then $\tderiv$ has the following shape:
\begin{center}
\begin{prooftree}[separation=1em, label separation = .2em]
					\hypo{\tderiv_1 \derives \typctxtwo \vdashpr \neu \hastype \tvar }
					\hypo{\tderiv_2 \derives \typctxthree \vdashpr \nf \hastype \mtype}
					\hypo{\tvartwo\mbox{ fresh, }\tderiv\disj\tderivtwo}
					\infer3[\scriptsize$\ruleApPr$]
					{\typctxtwo\isub\tvar{(\tarrow\mtype\tvartwo)} \uplus \typctxthree \vdashpr \neu \nf \hastype \tvartwo}
			\end{prooftree}
\end{center}
with $\typctx = \typctxtwo\isub\tvar{(\tarrow\mtype\tvartwo)} \uplus \typctxthree$. By \refpoint{dry_is_minimal-one}, $\tvar$ occurs exactly once in $\typctxtwo$. Then we have:
\begin{center}$\begin{array}{llllllllllllll}
\insize\tderiv & = & \insize{\tderiv_1} + \insize{\tderiv_2}+1
&=_{\ih}& \size\typctxtwo + \size\typctxthree +\size\mtype +1
\\
&=& \size\typctxtwo + \size\typctxthree +\size{\tarrow\mtype\tvartwo}
&=_{\refpointeq{dry_is_minimal-one}}& \size{\typctxtwo\isub\tvar{(\tarrow\mtype\tvartwo)}} + \size\typctxthree
\\
&=& \size{\typctxtwo\isub\tvar{(\tarrow\mtype\tvartwo)}} + \size\typctxthree + \size\tvartwo&&&\qedhere
\end{array}$\end{center}

\end{itemize}

\end{enumerate}
\end{proof}

%% file: proofs/dry_representation.tex
\begin{proof}
	By induction on $\tderiv$. Cases of the last rule: 
	\begin{itemize}
\item \emph{Rule $\ruleAx$}. Then $\tderiv$ is a $\ruleAx$ rule of conclusion $\var \hastype \mset\ltype \vdash \var \hastype \ltype$. The dry representation $\tderivtwo$ of $\tderiv$ is a $\ruleAx$ rule of conclusion $\var \hastype \mset\tvar \vdashpr \var \hastype \tvar$. The type substitution of the statement is $\isub\tvar\ltype$.

\item \emph{Rule $\ruleFun$}: it follows by the \ih

\item \emph{Rule $\ruleMany$}: it follows by the \ih Note that, by stability of dry derivations under renaming (\reflemma{dry-stabile-by-renaming}), we can assume that all the derivations $\tderivtwo_i$ given by the \ih are on disjoint supports, so that the constraint $\disj_{\iI}\tderivtwo_i$ for rule $\ruleManyPr$ is satisfied.

\item \emph{Rule $\ruleAp$}: then $\nf = \neu\nftwo$ and $\tderiv$ has the following shape:
\begin{center}$
\begin{prooftree}[separation=1em, label separation = .2em]
					\hypo{\tderiv_\neu \derives \typctx_\neu \vdash \neu \hastype \larrow{\mtype\!}{\!\ltype} }
					\hypo{\tderiv_{\nftwo} \derives \typctx_{\nftwo} \vdash \nftwo \hastype \mtype}
					\infer2[\scriptsize$\ruleAp$]
					{\typctx_\neu \uplus \typctx_{\nftwo} \vdash \neu\nftwo \hastype \ltype}
\end{prooftree}
$\end{center}
with $\typctx = \typctx_\neu \uplus \typctx_{\nftwo}$ and $\type = \ltype$. About the dry representation, by \ih there are dry derivations $\tderivtwo_\neu \derives \typctxtwo_\neu \vdashpr \neu \hastype \tvar$ and $\tderivtwo_{\nftwo} \derives \typctxtwo_{\nftwo} \vdashpr \nftwo \hastype \mtypetwo$ such that $\tderiv_\neu \tderiveq \tderivtwo_\neu$ and $\tderiv_{\nftwo} \tderiveq \tderivtwo_{\nftwo}$. By stability of dry derivations under renaming (\reflemma{dry-stabile-by-renaming}), we can assume that $\tderivtwo_\neu \disj \tderivtwo_\neu$. Then $\tderivtwo$ is obtained as follows:
\begin{center}$
\begin{prooftree}[separation=1em, label separation = .2em]
					\hypo{\tderivtwo_\neu \derives \typctxtwo_\neu \vdashpr \neu \hastype \tvar }
					\hypo{\tderivtwo_{\nftwo} \derives \typctxtwo_{\nftwo} \vdashpr \nftwo \hastype \mtypetwo}
					\hypo{\tvartwo \mbox{ fresh}, \tderivtwo_\neu \disj \tderivtwo_\neu}
					\infer3[\scriptsize$\ruleApPr$]
					{\typctxtwo_\neu\isub\tvar{(\tarrow\mtypetwo\tvartwo)} \uplus \typctxtwo_{\nftwo} \vdashpr \neu\nftwo \hastype \tvartwo}
\end{prooftree}
$\end{center}

About the type substitution, by \ih, there are substitutions $\sigma_\neu$ and $\sigma_{\nftwo}$ such that 
 $\typctxtwo_\neu\sigma_\neu = \typctx_\neu$ and $\tvar\sigma_\neu = \tarrow\mtype\ltype$, and 
$\typctxtwo_{\nftwo}\sigma_{\nftwo} = \typctx_{\nftwo}$ and $\mtypetwo\sigma_{\nftwo} = \mtype$. 
We can assume that $\dom{\sigma_\neu} = \dom{\typctxtwo_\neu}$ and $\dom{\sigma_{\nftwo}} = \dom{\typctxtwo_{\nftwo}}$, and we know that $\dom{\sigma_\neu}\cap \dom{\sigma_{\nftwo}}=\emptyset$. Define the substitution $\sigma$ as $\sigma_\neu$ on $\dom{\sigma_\neu}\setminus\set{\tvar}$, as $\sigma_{\nftwo}$ on $\dom{\sigma_{\nftwo}}$, and as $\isub\tvartwo\ltype$ on $\tvartwo$. Note that $\sigma_\neu(\tvar)= \tarrow\mtype\ltype$ and let $\sigma_\neu'$ be $\sigma_\neu$ without $\isub\tvar{(\tarrow\mtype\ltype)}$. Then:
\begin{center}$\begin{array}{llllllll}
(\typctxtwo_\neu\isub\tvar{(\tarrow\mtypetwo\tvartwo)} \uplus \typctxtwo_{\nftwo})\sigma & = & 
\multicolumn{5}{l}{\typctxtwo_\neu\sigma_\neu'\isub\tvar{(\tarrow{\mtypetwo\sigma_{\nftwo}}{\tvartwo\isub\tvartwo\ltype})} \uplus \typctxtwo_{\nftwo}\sigma_{\nftwo}}
\\
&=_\ih& 
\multicolumn{5}{l}{\typctxtwo_\neu\sigma_\neu'\isub\tvar{(\tarrow{\mtype}{\ltype})} \uplus \typctx_{\nftwo}}
\\
&=& \typctxtwo_\neu\sigma_\neu \uplus \typctx_{\nftwo}
&=_\ih& \typctx_\neu \uplus \typctx_{\nftwo}
&=& \typctx
\end{array}$\end{center}
and $\tvartwo\sigma=\tvartwo\isub\tvartwo\ltype=\ltype=\type$.\qedhere
\end{itemize}
\end{proof}

%% file: 07-Bounds_From_Composable_Types.tex
\section{Bounds From Composable Types}
\label{sect:bounds-comp-types}
In this section, we finally study bounds from composable types for leftmost evaluation and normal forms, which is the technical and neglected part of de Carvalho's work. To ease the study, we restrict to closed terms, so that type contexts disappear---de Carvalho does the same. There are however no issues in dealing with open terms.

\subparagraph*{Composable Types.} De Carvalho's idea is that, given two normal forms $\tm$ and $\tmtwo$, one can extract bounds for $\tm\tmtwo$ by looking only at the types of $\tm$ and $\tmtwo$---that is, at $\sem\tm$ and $\sem\tmtwo$---because a derivation for $\tm\tmtwo$ is just the application of a derivation for $\tm$ and one for $\tmtwo$.  We need to give a formal status to composable types, and we also need a notion of types that compose up to a type substitution.

\begin{definition}[Composable pairs]
Let $\tm$ and $\tmtwo$ be closed terms. 
\begin{itemize}
\item A \emph{(shrinking) composable pair} (of types) for $\tm$ and $\tmtwo$ is a pair $\compair= (\ltype, \mtype)$ such that $\ltype =\larrow{\mtype}{\ltypetwo} \in 
\sem\tm$,  $\mtype\in \sem\tmtwo$, and $\ltypetwo$ is right. The set 
of composable pairs of $\tm$ and $\tmtwo$ is noted $ \xcompairs\tm\tmtwo$.
\item A \emph{(shrinking) composable pair up to substitution} for $\tm$ and $\tmtwo$ is a pair $\compair= (\ltype, \mtype)$ such that there exists a type substitution $\sigma$ such that $(\ltype\sigma, \mtype\sigma)\in \xcompairs\tm\tmtwo$. The set 
of composable pairs up to substitution of $\tm$ and $\tmtwo$ is noted as $\xsubcompairs\tm\tmtwo$.
\end{itemize}
\end{definition}
Note that if $(\larrow{\mtype}{\ltypetwo},\mtype)\in \xcompairs\tm\tmtwo$ only $\ltypetwo$ is required to be a right shrinking type (as it is the only type in the judgement for $\tm\tmtwo$ after composition), while $\larrow{\mtype}{\ltypetwo}$ might very well not be a right shrinking type (if $\mtype$ is not left). The constraint that $\ltypetwo$ is right in the definition of $\xcompairs\tm\tmtwo$ ensures the following property.
\begin{lemma}[Normalization and composable types]
\label{l:norm-and-comppairs-nonempty}
Let $\tm$ and $\tmtwo$ be closed terms. Then $\tm\tmtwo$ $\tolo$-normalizes if and only if $\xcompairs\tm\tmtwo \neq \emptyset$. 
\end{lemma}

\input{\proofspath/leftmost/normalization_composable_bounds}

\subparagraph{Normal Forms and Composable Types.} To warm up, we first show how to bound the size of normal forms from composable pairs. 
\begin{theorem}[Normal form bounds from composable types]
	\label{thm:nf-bounds-3}
Let $\tm$ and $\tmtwo$ be closed terms such that there is a normalizing evaluation $\deriv:\tm\tmtwo \tolo^*\nf$.
\begin{enumerate}
\item \emph{Lax bounds}: $\sksize{\nf} \leq \size\ltype$ for every composable pair $(\tarrow\mtype\ltype, \mtype)\in \xcompairs\tm\tmtwo$. 
\item \emph{Exact bounds from special pairs}: moreover, there exists a composable pair $(\tarrow\mtype\ltype, \mtype)\in \xcompairs\tm\tmtwo$ such that $\sksize{\nf} =\size\ltype$.
\end{enumerate}
\end{theorem}
\input{\proofspath/leftmost/nf_composable_bounds}

\subparagraph{Lax Evaluation Bounds from Composable Types.} Now, we study how to additionally extract (bounds on) the number of leftmost step from a composable pair. Obtaining lax bounds is easy. The idea is that the composed type bounds the size of a derivation for $\tm\tmtwo$, which in turns provides bounds about $\tm\tmtwo$, as shown in \refsect{bounds-derivations}. We also show that even composable pairs up to substitution yield bounds.

\begin{theorem}[Lax bounds from composable types]
	\label{thm:lax-bounds-3}
Let $\nf$ and $\nftwo$ be closed normal terms such that $\deriv:\nf\nftwo \tolo^* \nfthree$ with $\nfthree$ normal. Then:
\begin{enumerate}
\item \emph{Lax bounds and types}: $2\size{\deriv} + \sksize{\nfthree} \leq \size\ltype+\size\mtype+1$ for every composable pair $(\ltype, \mtype)\in \xcompairs\nf\nftwo$.
\item \emph{Lax bounds and types, up to substitutions}: $2\size{\deriv} + \sksize{\nfthree} \leq \size\ltype+\size\mtype+1$ for every composable pair up to substitution $(\ltype, \mtype)\in \xsubcompairs\nf\nftwo$.
\end{enumerate}
\end{theorem}

\input{\proofspath/leftmost/lax_composable_bounds}

Note that the hypotheses for bounding evaluation lengths are stronger than for bounding normal forms (\refthm{nf-bounds-3}), as the two applied terms $\nf$ and $\nftwo$ are required to be normal. If the two composed terms are not normal, then there is no way to measure the extra steps to their normal forms using their types, because types are invariant by reduction. The stronger hypotheses limit the scope of the result: if $\tm\defeq \la\var\var\Omega\nf$ with $\nf$ normal and $\tmtwo\defeq \la\vartwo\la\varthree\varthree$ then $\tm\tmtwo \tolo^3 \nf$ and \refthm{nf-bounds-3} can be applied, while \refthm{lax-bounds-3} cannot, because of the diverging $\Omega$ sub-term in $\tm$.

\subparagraph{Exact Bounds from Composable Types.} Now, the tricky point is how to obtain \emph{exact} bounds. The problem is that for the application of two normal terms $\nf$ and $\nftwo$, the minimal composable pair in general does \emph{not} provide exact bounds, and---dually---minimal types for $\nf$ and $\nftwo$ do \emph{not} compose. The following example pinpoints the subtleties.

\input{figure-derivation_with_minimal_types}
\subparagraph*{Key Example.} 
Consider the unitary shrinking derivations $\tderiv_\delta$ and $\tderiv_\Id$ of minimal types for  $\delta \defeq \la\var\var\var$ and for $\Id \defeq \la\vartwo\vartwo$:
\begin{center}
\begin{tabular}{c@{\hspace{1cm}}|@{\hspace{1cm}}c}
			$\tderiv_\delta =
			\begin{prooftree}[separation=.7em]
			\hypo{}
			\infer1[\footnotesize$\ruleAx$]{\var\hastype \mset{\larrow{\mset\tvarthree}{\tvarthree}} \vdash \var\hastype \larrow{\mset\tvarthree}{\tvarthree} }
			\hypo{}
			\infer1[\footnotesize$\ruleAx$]{\var\hastype \mset\tvarthree \vdash \var\hastype \tvarthree }
			\infer2[\footnotesize$\ruleAp$]{\var\hastype \mset{\larrow{\mset\tvarthree}{\tvarthree}, \tvarthree} \vdash \var\var \hastype \tvarthree}
			\infer1[\footnotesize$\ruleFun$]{\vdash \la{\var}\var\var \hastype \larrow{\mset{\larrow{\mset\tvarthree}{\tvarthree}, \tvarthree}}{\tvarthree}}
			\end{prooftree}$
&
			$\tderiv_\Id = 
			\begin{prooftree}[separation=1em]
			\hypo{}
			\infer1[\footnotesize$\ruleAx$]{\vartwo\hastype \mset\tvarfour \vdash \vartwo\hastype \tvarfour }
			\infer1[\footnotesize$\ruleFun$]{\vdash \la{\vartwo}\vartwo \hastype \larrow{\mset\tvarfour}{\tvarfour}}
			\end{prooftree}$
\end{tabular}
\end{center}
			Note that, pleasantly, $\insize{\tderiv_\delta} = 2 = \sizetyp{\larrow{\mset{\larrow{\mset\tvarthree}{\tvarthree}, \tvarthree}}{\tvarthree}} = \insize{\delta}$, and $\insize{\tderiv_\Id} = 1 = \sizetyp{\larrow{\mset\tvarfour}{\tvarfour}} = \insize{\Id}$.
Now, consider the application $\delta\Id$. Note that, unfortunately, the two obtained minimal types do \emph{not} compose, and not just because they use different type variables: identifying $\tvarthree$ and $\tvarfour$ would not be enough, one actually needs to identify $\tvarthree$ with $\tvarfour^2 \defeq \larrow{\mset\tvarfour}{\tvarfour}$. The unitary shrinking derivation $\tderivtwo_{\delta\Id}$ for $\delta\Id$ with minimal types (which provides exact information for $\delta\Id$) obtained in this way is in \reffig{composed-deriv}. Note that its sub-derivations $\tderivtwo_\delta$ and $\tderivtwo_\Id$ for $\delta$ and $\Id$ do \emph{not} derive minimal types. The derivation  $\tderivtwo_{\delta\Id}$ indeed  is obtained by composing:
\begin{enumerate}
\item The variant $\tderivtwo_\delta$ of $\tderiv_\delta$ which has the same exact structure of $\tderiv_\delta$ and where every occurrence of $\tvarthree$ has been replaced by $\tvarfour^2$, obtaining the type $\larrow{\mset{\mset{\larrow{\mset{\tvarfour^2}}{\tvarfour^2}, {\tvarfour^2}}}}{\tvarfour^2}$,
\item With $\tderivtwo_\Id$, which puts together two derivations for $\Id$, one being $\tderiv_\Id$ (of type $\tvarfour^2$), and one being the variant $\tderiv_\Id'$ of $\tderiv_\Id$ (of type $\larrow{\mset{\tvarfour^2}}{\tvarfour^2}$) where $\tvarfour$ has been replaced by $\tvarfour^2$.
\end{enumerate}
Note that there is a \emph{gap} between:
\begin{itemize}
\item The length of the evaluation $\deriv: \delta\Id \tolo \Id\Id \tolo \Id$, that takes 2 steps, plus the size of the normal form $\insize{\Id}=1$, so that $2\size\deriv + \insize{\Id}=5$, and
\item The size of the composable pair $\compair = (\larrow{\mset{\larrow{\mset{\tvarthree^2}}{\tvarthree^2}, {\tvarthree^2}}}{\tvarthree^2}, \mset{\larrow{\mset{\tvarthree^2}}{\tvarthree^2}, {\tvarthree^2}})$, which is $10$.
\end{itemize}
The point is that  the types derived by $\tderivtwo_\delta$ and $\tderivtwo_\Id$ are \emph{not} minimal, so their sizes are bigger than $\insize{\tderivtwo_\delta}$ and $\insize{\tderivtwo_\Id}$, and do not provide exact bounds for $\delta\Id$. For instance, the size of $\larrow{\mset{\larrow{\mset{\tvarfour^2}}{\tvarfour^2}, {\tvarfour^2}}}{\tvarfour^2}$, which is the type of $\tderivtwo_\delta$, is 6, while $\insize{\tderivtwo_\delta}=2$---this is an instance of the mentioned gap. 
Summing up, \emph{minimal types do not compose}, and \emph{composable types do not give exact bounds}. 

\subparagraph*{Out of the Impasse.}De Carvalho solves this \emph{cul-de-sac} using composable pairs \emph{up to substitution}. With respect to our example, he considers the composable pair $\compair$ given by $\tderivtwo$, but computes the bound using the pair $\compairtwo = (\larrow{\mset{\larrow{\mset\tvarthree}{\tvartwo}, \tvarthree}}{\tvartwo}, \mset{\larrow{\mset\tvar}{\tvar},\larrow{\mset{\tvar'}}{\tvar'}})$, which is minimal and non-composable. It is obtained by collecting the types of the dry version of $\tderivtwo_\delta$ (of type $\larrow{\mset{\larrow{\mset\tvarthree}{\tvartwo}, \tvarthree}}{\tvartwo}$) and the dry version of $\tderivtwo_\Id$ (typing $\Id$ twice thus having type $\mset{\larrow{\mset\tvar}{\tvar},\larrow{\mset{\tvar'}}{\tvar'}}$). The last bit is noting that $\compairtwo$ is composable \emph{up to the substitution} $\sigma \defeq \isub\tvarthree{\tvarfour^2}\isub\tvartwo{\tvarfour^2}\isub\tvar{\tvarfour^2}\isub{\tvar'}{\tvarfour}$, since $\compairtwo\sigma=\compair$.

Roughly, for minimal types to compose, they usually have to be expanded, as we have done in the example when substituting $\tvarthree$ with $\tvarfour^2$. Such an expansion introduces some noise in the measures, so that even minimal composable pairs might not provide exact bounds. De Carvalho's trick is to reverse the expansion, considering composable pairs up to substitution. Our notion of dry derivation makes the expansion reversal technically clean.

\subparagraph{Main Result.} We can now prove the main result of the paper, namely de Carvalho's exact bounds from composable types. 
\begin{theorem}[Exact bounds from composable types]
	\label{thm:new-exact-bounds-3}
	Let $\nf$ and $\nftwo$ be normal. 
	If $\deriv:\nf\nftwo \tolo^* \nfthree$ and $\nfthree$ is normal. Then:
	\begin{enumerate}
	\item \emph{Exact bounds}: there 
	exist $\ltype\in \sem{\nf}$ and $\mtype \in \sem{\nftwo}$ such that $2\size{\deriv} + \sksize\nfthree = 
	\size\ltype+\size\mtype+1$, and $\ltype$ and $\mtype$ are obtained by drying the composable pair induced by a unitary shrinking derivation for $\nf\nftwo$. 
	\item \emph{From types composable up to substitution}: moreover, $(\ltype,\mtype)$ are composable up to substitution, that is, $(\ltype,\mtype)\in\xsubcompairs\nf\nftwo$.
	\end{enumerate}
\end{theorem}

\input{\proofspath/leftmost/composable_bounds}

%
%


%% file: proofs/leftmost/normalization_composable_bounds.tex
\begin{proof}
If $\tm\tmtwo$ normalizes then by shrinking completeness (\refthm{shrinking-characterization}) there exists a shrinking 
derivation $\tderiv\vdash\tm\tmtwo \hastype \ltype$ with $\ltype$ right. The last rule of $\tderiv$ is $\ruleAp$ and its premises give a composable pair $(\tarrow\mtype\ltype,\mtype)$ for $\tm$ and $\tmtwo$, for some $\mtype$. Therefore, $\xcompairs\tm\tmtwo \supset \set{(\tarrow\mtype\ltype,\mtype)}\neq\emptyset$.

 Vice versa, if $\xcompairs\tm\tmtwo\neq\emptyset$ then every composable pair 
induces a shrinking derivation for $\tm\tmtwo$, by connecting the two derivations producing the composable pair via rule $\ruleAp$. Thus, $\tm\tmtwo$ is typable. By shrinking correctness (\refthm{shrinking-characterization}), $\tm\tmtwo$ is $\tolo$-normalizing.
\end{proof}

%% file: proofs/leftmost/nf_composable_bounds.tex
\begin{proof}
\hfill
\begin{enumerate}

\item Let $(\larrow{\mtype}{\ltype}, \mtype) \in \xcompairs\tm\tmtwo$ be a composable pair. Then, there are derivations $\concl{\tderiv_\tm}{}{\tm}{\larrow{\mtype}{\ltype}}$ and $\concl{\tderiv_\tmtwo}{}{\tmtwo}{\mtype}$. We compose them as a derivation $\tderiv$ for $\tm\tmtwo$ via  rule $\ruleAp$:
\begin{center}$
		\tderiv \defeq 
		\begin{prooftree}
		\hypo{\tderiv_{\tm}\derives\vdash \tm \hastype \larrow{\mtype}{\ltype}}
		\hypo{\tderiv_{\tmtwo} \derives \vdash \tmtwo \hastype\mtype}
		\infer2[\footnotesize$\ruleApp$]{\tyjp{}{\tm\tmtwo}{}{\ltype}}
		\end{prooftree}$
		\end{center}
		Note that the definition of composable pair guarantees that $\ltype$ is right (shrinking), so that $\tderiv$ is shrinking.
By shrinking correctness (\refthm{shrinking-characterization}), there is a derivation $\vdash \nf\hastype \ltype$. Since shrinking types bound the size of normal forms (\refprop{shrinking-nfs-bounds-from-types}), $\sksize\nf \leq \size\ltype$. 

\item By \refpropp{shrinking-nfs-bounds-from-types}{exists}, there exists a unitary shrinking derivation  $\tderivtwo\derives \vdash \nf\hastype\ltype$ for $\nf$ such that $\size\ltype= \sksize\nf$.
 Pulling back the final judgement of $\tderivtwo$ using subject expansion (\refpropp{qual-subject}{expansion}), we obtain a derivation $\tderivthree \vdash \tm\tmtwo\hastype \ltype$.
 The last rule of $\tderivthree$ is $\ruleAp$ and its premises give a composable pair $(\tarrow\mtype\ltype,\mtype)$ for $\tm$ and $\tmtwo$, for some $\mtype$.\qedhere
\end{enumerate}
\end{proof}

%% file: proofs/leftmost/lax_composable_bounds.tex
\begin{proof} 
\hfill
\begin{enumerate}
\item Let $(\ltype, \mtype) \in \xcompairs\nf\nftwo$ be a composable pair, which implies $\ltype= \larrow{\mtype}{\ltypetwo}$ for some right $\ltypetwo$. Then, there are two derivations $\concl{\tderiv_\nf}{}{\nf}{\larrow{\mtype}{\ltypetwo}}$ and $\concl{\tderiv_{\nftwo}}{}{\nftwo}{\mtype}$. We compose them via a $\ruleAp$ rule into a derivation $\tderiv$ for $\nf\nftwo$:
\begin{center}
		$\tderiv \defeq 
		\begin{prooftree}
		\hypo{\tderiv_{\nf}\derives\vdash \nf \hastype \larrow{\mtype}{\ltypetwo}}
		\hypo{\tderiv_{\nftwo} \derives \vdash \nftwo \hastype\mtype}
		\infer2[\footnotesize$\ruleApp$]{\tyjp{}{\nf\nftwo}{}{\ltypetwo}}
		\end{prooftree}$
		\end{center}
By definition of composable pair, $\ltypetwo$ is right shrinking, so that $\tderiv$ is shrinking.
By shrinking correctness (\refthm{shrinking-characterization}), $2\size{\deriv} + \sksize\nfthree \leq \insize\tderiv = \insize{\tderiv_\nf} + \insize{\tderiv_{\nftwo}} + 1$. Now, since types bound the size of the derivation for normal terms (\refprop{types-bound-normal-derivations}),  we obtain $\insize{\tderiv_\nf} \leq \size{\larrow{\mtype}{\ltypetwo}}$ and $\insize{\tderiv_{\nftwo}} \leq \size\mtype$. Therefore,  $2\size{\deriv} + \sksize\nfthree \leq \insize\tderiv  \leq \size{\larrow{\mtype}{\ltypetwo}}+\size\mtype+1$, as required.

\item Let $(\ltype, \mtype) \in \xsubcompairs\nf\nftwo$ be a composable pair up to substitution, which implies that there exist a type substitution $\sigma$ such that $\ltype\sigma= \larrow{\mtypetwo}{\ltypetwo}$, $\mtype\sigma = \mtypetwo$ for  some right $\ltypetwo$ and some $\mtypetwo$. Then, there are two  derivations $\concl{\tderiv_\nf}{}{\nf}{\ltype}$ and $\concl{\tderiv_{\nftwo}}{}{\nftwo}{\mtype}$. By \reflemmap{dry_is_standard}{sub}, applying $\sigma$ to $\tderiv_\nf$ and $\tderiv_{\nftwo}$ we obtain two derivations $\concl{\tderivtwo_\nf}{}{\nf}{\larrow{\mtypetwo}{\ltypetwo}}$ and $\concl{\tderivtwo_{\nftwo}}{}{\nftwo}{\mtypetwo}$ such that $\tderiv_\nf \tderiveq \tderivtwo_\nf$ and $\tderiv_{\nftwo} \tderiveq \tderivtwo_{\nftwo}$. We compose them via a $\ruleAp$ rule into a derivation $\tderivtwo$ for $\nf\nftwo$:
\begin{center}$
		\tderivtwo \defeq 
		\begin{prooftree}
		\hypo{\tderivtwo_{\nf}\derives\vdash \nf \hastype \larrow{\mtypetwo}{\ltypetwo}}
		\hypo{\tderivtwo_{\nftwo} \derives \vdash \nftwo \hastype\mtypetwo}
		\infer2[\footnotesize$\ruleApp$]{\tyjp{}{\nf\nftwo}{}{\ltypetwo}}
		\end{prooftree}
$\end{center}
Since $\ltypetwo$ is right, $\tderivtwo$ is shrinking.
By shrinking correctness (\refthm{shrinking-characterization}), $2\size{\deriv} + \sksize\nfthree \leq \insize\tderivtwo = \insize{\tderivtwo_\nf} + \insize{\tderivtwo_{\nftwo}} + 1$. By the skeletal invariants (\reflemma{skel-equiv-properties}), we obtain $\insize{\tderivtwo_\nf} + \insize{\tderivtwo_{\nftwo}} + 1= \insize{\tderiv_\nf} + \insize{\tderiv_{\nftwo}} + 1$. Since types bound the size of the derivation for normal terms (\refprop{types-bound-normal-derivations}),  $\insize{\tderiv_\nf} \leq \size{\ltype}$ and $\insize{\tderiv_{\nftwo}} \leq \size\mtype$. Therefore,  $2\size{\deriv} + \sksize\nfthree \leq \size{\ltype}+\size\mtype+1$, as required.
\qedhere
\end{enumerate}
\end{proof}

%% file: figure-derivation_with_minimal_types.tex
\begin{figure*}[t!]
\centering
\scriptsize
			\begin{prooftree}[separation=1em]
			\hypo{}
			\infer1[\footnotesize$\ruleAx$]{\var\hastype \mset{\larrow{\mset{\tvarfour^2}}{\tvarfour^2}} \vdash \var\hastype \larrow{\mset{\tvarfour^2}}{\tvarfour^2} }
			\hypo{}
			\infer1[\footnotesize$\ruleAx$]{\var\hastype \mset{\tvarfour^2} \vdash \var\hastype {\tvarfour^2} }
			\infer1[\footnotesize$\ruleManyVar$]{\var\hastype \mset{\tvarfour^2} \vdash \var\hastype \mset{\tvarfour^2} }	
			\infer2[\footnotesize$\ruleAp$]{\var\hastype \mset{\larrow{\mset{\tvarfour^2}}{\tvarfour^2}, {\tvarfour^2}} \vdash \var\var \hastype \tvarfour^2}
			\infer1[\footnotesize$\ruleFun$]{\tderivtwo_\delta \derives \vdash \la{\var}\var\var \hastype \larrow{\mset{\larrow{\mset{\tvarfour^2}}{\tvarfour^2}, {\tvarfour^2}}}{\tvarfour^2}}			
						\hypo{}
			\infer1[\footnotesize$\ruleAx$]{\vartwo\hastype \mset{\tvarfour^2} \vdash \vartwo\hastype {\tvarfour^2} }
			\infer1[\footnotesize$\ruleFun$]{\vdash \la{\vartwo}\vartwo \hastype \larrow{\mset{\tvarfour^2}}{\tvarfour^2}}			
			
			\hypo{}
			\infer1[\footnotesize$\ruleAx$]{\vartwo\hastype \mset\tvarfour \vdash \vartwo\hastype \tvarfour }
			\infer1[\footnotesize$\ruleFun$]{\vdash \la{\vartwo}\vartwo \hastype \larrow{\mset\tvarfour}{\tvarfour}}
			\infer2[\footnotesize$\ruleManyVal$]{\tderivtwo_\Id \derives \vdash \la{\vartwo}\vartwo \hastype \mset{\larrow{\mset{\tvarfour^2}}{\tvarfour^2}, \tvarfour^2}}
			\infer2[\footnotesize$\ruleAp$]{\vdash \delta\Id \hastype \tvarfour^2}
			\end{prooftree}

\caption{Unitary shrinking derivation $\tderivtwo_{\delta\Id}$ of minimal type for $\delta\Id$, where $\tvarfour^2 \defeq \larrow{\mset\tvarfour}{\tvarfour}$.}
\label{fig:composed-deriv}
\end{figure*}

%% file: proofs/leftmost/composable_bounds.tex
\begin{proof}
\begin{enumerate}
\item 
	By unitary shrinking completeness (\refthm{unitary-shrinking-characterization}), there is a unitary shrinking derivation $\derive{\tderiv}{\nf\nftwo}$ which, by unitary shrinking correctness, satisfies $2\size{\deriv} + \sksize{\nfthree} = \insize{\tderiv}$. 
	The last rule of $\tderiv$ is an $\ruleAp$ rule, that is, $\tderiv$ has the following shape:
	\begin{center}
		\begin{prooftree}
		\hypo{\tderiv_{\nf}\derives\vdash \nf \hastype \larrow{\mtypetwo}{\ltypetwo}}
		\hypo{\tderiv_{\nftwo} \derives \vdash \nftwo \hastype\mtypetwo}
		\infer2[\footnotesize$\ruleApp$]{\tyjp{}{\nf\nftwo}{}{\ltypetwo}}
		\end{prooftree}
\end{center}
With $\ltypetwo$ right linear type. Note that $\insize{\tderiv} = \insize{\tderiv_\nf} + \insize{\tderiv_{\nftwo}} + 1$. 
		By dry representation (\refthmp{dry_representation}{core}), there are dry derivations $\concl{\tderivtwo_\nf}{\,}{\nf}{\ltype}$ and $\concl{\tderivtwo_{\nftwo}}{\,}{\nftwo}{\mtype}$ such that $\tderiv_\nf \tderiveq \tderivtwo_\nf$ and $\tderiv_{\nftwo} \tderiveq \tderivtwo_{\nftwo}$. By the properties of skeletal invariants (\reflemma{skel-equiv-properties}), $\insize{\tderiv_\nf} = \insize{\tderivtwo_\nf}$ and $\insize{\tderiv_{\nftwo}} = \insize{\tderivtwo_{\nftwo}}$. By the fact that dry derivations have minimal types (\refprop{dry_is_minimal}), $\insize{\tderivtwo_\nf}= \size\ltype$ and $\insize{\tderivtwo_{\nftwo}}= \size\mtype$. Putting it all together, we obtain: 
		\begin{center}$\begin{array}{llllllllll}
		2\size{\deriv} + \sksize{\nfthree} & =_{\refthmeq{unitary-shrinking-characterization}} &\insize{\tderiv_\nf} + \insize{\tderiv_{\nftwo}} + 1 
		\\
		&=_{\reflemmaeq{skel-equiv-properties}} &\insize{\tderivtwo_\nf} + \insize{\tderivtwo_{\nftwo}} + 1 
		 
		& =_{\refpropeq{dry_is_minimal}} & \size\ltype+\size\mtype+1
		\end{array}$\end{center}
		
		\item By the \emph{type substitution} part of the dry representation theorem (\refthmp{dry_representation}{sub}), there exist substitutions $\sigma_\nf$ and $\sigma_{\nftwo}$ such that $\ltype\sigma_\nf = \larrow{\mtypetwo}{\ltypetwo}$ and $\mtype\sigma_{\nftwo} = \mtypetwo$. 
By stability of dry derivations under renaming (\reflemma{dry-stabile-by-renaming}), we can assume that the supports of $\tderivtwo_\nf$ and $\tderivtwo_{\nftwo}$ are disjoint, so that the domains of $\sigma_\nf$ and $\sigma_{\nftwo}$ are disjoint, allowing us to define $\sigma$ as $\sigma_\nf \cup \sigma_{\nftwo}$, for which $\ltype\sigma = \ltype\sigma_\nf = \larrow{\mtypetwo}{\ltypetwo}$ and $\mtype\sigma = \mtype\sigma_{\nftwo} = \mtypetwo$. Finally, note that  $\ltypetwo$ is right by hypothesis (because $\tderiv$ is shrinking), so that $(\ltype,\mtype)\in\xsubcompairs\nf\nftwo$. \qedhere
		\end{enumerate}
\end{proof}

%% file: 08-Head_Case.tex
\section{The Less Satisfying Head Case}
\label{sect:head-case}
We conclude our study by adapting the bounds from composable types to the case of head reduction. The study is slightly different in that we omit the study of type substitutions and dry derivations. We do so to show that one can obtain the first part of the de Carvalho's result---which in our opinion is the important one---by resting only on the simpler \emph{size representation theorem} of \refsect{dissecting}, with no need of bothering about countably many type variables and dry derivations. 

We give only the proof of the main theorem, as the other ones are variants of those in the previous section. They can be found \techreport{in the Appendix.}\paper{in the technical report \cite{accattoli2023semantic}.}

\subparagraph{The Head Case.} Let $\hcompairs\tm\tmtwo$ and $\hsubcompairs\tm\tmtwo$ be the analogous sets of $\xcompairs\tm\tmtwo$ and $\xsubcompairs\tm\tmtwo$ but without asking that the composed type is right shrinking, which is not needed for characterizing head termination.

\begin{lemma}[Head normalization and composable types]
\label{l:norm-and-comppairs-nonempty-head}
\techreport{\NoteProof{lappendix:norm-and-comppairs-nonempty-head}}
Let $\tm$ and $\tmtwo$ be closed terms. Then $\tm\tmtwo$ $\toh$-normalizes if and only if $\hcompairs\tm\tmtwo \neq \emptyset$. 
\end{lemma}

The next theorem adapts lax bounds.

\begin{theorem}[Lax bounds for head reduction from composable types]
	\label{thm:lax-bounds-3-head}
\techreport{	\NoteProof{thmappendix:lax-bounds-3-head}}
Let $\nf$ and $\nftwo$ be closed normal terms such that $\deriv:\nf\nftwo \tolo^* \hnf$ with $\hnf$ head normal. Then:

\begin{enumerate}
\item \emph{Lax bounds and types}: $2\size{\deriv} + \hdsize{\hnf} \leq \size\ltype+\size\mtype+1$ for every composable pair $(\ltype, \mtype)\in \hcompairs\nf\nftwo$.
\item \emph{Lax bounds and types, up to substitutions}: $2\size{\deriv} + \hdsize{\hnf} \leq \size\ltype+\size\mtype+1$ for every composable pair up to substitution $(\ltype, \mtype)\in \hsubcompairs\nf\nftwo$.
\end{enumerate}
\end{theorem}

\begin{theorem}[Exact bounds for head reduction from composable types]
	\label{thm:new-exact-bounds-3-head}
	Let $\nf$ and $\nftwo$ be normal and such that $\deriv:\nf\nftwo \toh^* \hnf$ with $\hnf$ head normal. Then there 
	exist $\ltype\in \sem{\nf}$ and $\mtype \in \sem{\nftwo}$ such that $2\size{\deriv} + \hdsize\hnf = 
	\size\ltype+\size\mtype+1$. 
\end{theorem}

\input{\proofspath/head/composable_bounds}

Note that the hypotheses \emph{$\nf$ and $\nftwo$ are normal} is not a typo, we do mean \emph{normal}, not \emph{head normal}. The stronger hypotheses are required because the evaluation $\deriv$ leading to $\hnf$ might involve arbitrary sub-terms of $\nf$ and $\nftwo$, not just their heads. For instance if $\nf \defeq \la\var\var\tm$ and $\nftwo\defeq \Id$ then $\nf\nftwo \toh \Id\tm \toh \tm$, and $\tm$ is not a head sub-term of $\nf$.

This is where the results are less satisfying. Having to assume that $\nf$ and $\nftwo$ are normal might be acceptable for leftmost reduction but it limits considerably the value of de Carvalho's analysis in the head case, which is the case naturally corresponding to relational semantics. Indeed, there are many head normalizing terms that have no normal form---the paradigmatic example begin given by \emph{fix-point operators}---and about which \refthm{new-exact-bounds-3-head} does not say anything.

%% file: proofs/head/composable_bounds.tex
\begin{proof}
	By head completeness (\refthm{head-characterization}), there is a derivation $\derive{\tderiv}{\nf\nftwo}$  satisfying $2\size{\deriv} + \hdsize{\hnf} = \insize{\tderiv}$. 
	The last rule of $\tderiv$ is an $\ruleAp$ rule, that is, $\tderiv$ has the following shape:
	\begin{center}
		\begin{prooftree}
		\hypo{\tderiv_{\nf}\derives\vdash \nf \hastype \larrow{\mtypetwo}{\ltypetwo}}
		\hypo{\tderiv_{\nftwo} \derives \vdash \nftwo \hastype\mtypetwo}
		\infer2[\footnotesize$\ruleApp$]{\tyjp{}{\nf\nftwo}{}{\ltypetwo}}
		\end{prooftree}
\end{center}
Note that $\insize{\tderiv} = \insize{\tderiv_\nf} + \insize{\tderiv_{\nftwo}} + 1$. 
		By size representation (\refthm{size-representation}), there are $\concl{\tderivtwo_\nf}{\,}{\nf}{\ltype}$ and $\concl{\tderivtwo_{\nftwo}}{\,}{\nftwo}{\mtype}$ such that $\tderiv_\nf \tderiveq \tderivtwo_\nf$ and $\insize{\tderivtwo_\nf}= \size\ltype$, and
		 $\tderiv_{\nftwo} \tderiveq \tderivtwo_{\nftwo}$ and $\insize{\tderivtwo_{\nftwo}}= \size\mtype$.

		By the properties of skeletal invariants (\reflemma{skel-equiv-properties}), $\insize{\tderiv_\nf} = \insize{\tderivtwo_\nf}$ and $\insize{\tderiv_{\nftwo}} = \insize{\tderivtwo_{\nftwo}}$. Putting it all together, we obtain: 
		\begin{center}$\begin{array}{llllllllll}
		2\size{\deriv} + \hdsize{\hnf} & =_{\refthmeq{head-characterization}} &\insize{\tderiv_\nf} + \insize{\tderiv_{\nftwo}} + 1 
		\\
		&=_{\reflemmaeq{skel-equiv-properties}} &\insize{\tderivtwo_\nf} + \insize{\tderivtwo_{\nftwo}} + 1 
		
		& =_{\refthmeq{size-representation}} & \size\ltype+\size\mtype+1. \qedhere
		\end{array}$\end{center}
%
\end{proof}

%% file: 10-Conclusions.tex
\section{Conclusions}
\label{sect:conclusions}
In 2007, de Carvalho developed a sharp quantitative analysis of the $\l$-calculus using multi types and the relational model. His study has been influential, leading to a recent new wave of studies in the $\l$-calculus halfway between operational and denotational semantics. Only the first and simpler half of de Carvalho's results, however, has really permeated the community. The second more technical---and probably even more interesting---part, which lifts the quantitative analysis to the relational model, has instead been ignored by the recent literature.  This paper dissects it and revisits it, pointing out the underlying subtleties and clarifying the concepts and tools for its proof.

A preliminary version of this work led to the adaption of de Carvalho's compositional bounds to call-by-value, which can be found in the technical report \cite{DBLP:journals/corr/abs-2104-13979} by Accattoli et al. Hopefully, further adaptations to other $\l$-calculi will be developed. 

About future work, a sharper study for the head case should be developed, as to avoid the normal form hypotheses. The main weakness of de Carvalho's results, indeed, is that they really work only for strong reduction, at present. As we hinted at in the introduction, probably a finer understanding of the external \emph{vs} internal behaviour of terms is needed.

%% file: 99-00-Appendix.tex
\section*{Proofs Appendix}
\label{sect:app-index}
The original material in the paper is contained in \Cref{sect:dissecting}, \Cref{sect:bounds-comp-types}, and \Cref{sect:head-case}. The proof appendix contains the few proofs omitted from \Cref{sect:dissecting} and  \Cref{sect:head-case}. 

\input{99-06-Dry_Derivations}


\input{99-08-Head_Bounds_From_Composable_Types}

%% file: 99-06-Dry_Derivations.tex
\section{Proofs of \Cref{sect:dissecting} (Dissecting Bounds From Types via Skeletons and Dry Judgements)}

\subparagraph{Removing Substitutions.} Here we prove the size representation theorem. For the induction to go through, we need to add two further statements, proved by mutual induction with the main one, which is the second point of the next theorem.

\begin{theorem}[Size representation]
\label{thmappendix:size-representation}
\NoteState{thm:size-representation}
Let $\tm$ be a term and $\tderiv \derives \typctx \vdash \tm \hastype \type$ be a derivation.
\begin{enumerate}
\item If $\tm$ is neutral and $\type$ is a linear type $\ltype$, then for every $1$-type $\ltypetwo$ there exists a derivation $\tderivtwo \derives \typctxtwo \vdashun \tm \hastype \ltypetwo$ such that $\tderivtwo \tderiveq \tderiv$ and $\insize\tderivtwo = \sizectx\typctxtwo - \size\ltypetwo$.

\item If $\tm$ is normal and $\type$ is a linear type $\ltype$, then there exists a derivation $\tderivtwo \derives \typctxtwo \vdashun \tm \hastype \ltypetwo$ such that $\tderivtwo \tderiveq \tderiv$ and $\insize\tderivtwo = \sizectx\typctxtwo + \size\ltypetwo$.

\item If $\tm$ is normal and $\type$ is a multi type $\mtype$, then there exists a derivation $\tderivtwo \derives \typctxtwo \vdashun \tm \hastype \mtypetwo$ such that $\tderivtwo \tderiveq \tderiv$ and $\insize\tderivtwo = \sizectx\typctxtwo + \size\mtypetwo$.
\end{enumerate}
\end{theorem}

\input{\proofspath/size_representation}

%% file: proofs/size_representation.tex
\begin{proof}
	By mutual induction on the definition of neutral and normal terms, followed by an induction on the type derivation $\tderiv$.
	\begin{enumerate}
	\item \emph{$\tm$ is a neutral term and $\type$ is a linear type $\ltype$.} Cases of the last rule:
	\begin{itemize}		
		\item \emph{Rule $\ruleAx$}, that is, $\tm = \var$. Then:
\begin{center}$
		\tderiv = 
		\begin{prooftree}
		\infer0[\footnotesize$\ruleAx$]{\tyjp{}{\var}{\var \hastype \mset{\ltype}}{\ltype}}
		\end{prooftree}
$\end{center}
		where $\typctx = \var \hastype \mset\ltype$. Then, for every $1$-type $\ltypetwo$, the derivation $\tderivtwo$ of the statement is defined as follows:
\begin{center}$
		\tderivtwo \defeq 
		\begin{prooftree}
		\infer0[\footnotesize$\ruleAx$]{\tyjp{}{\var}{\var \hastype \mset{\ltypetwo}}{\ltypetwo}}
		\end{prooftree}
$\end{center}
		with $\typctxtwo \defeq \mset{\ltypetwo}$. Note that $\tderiv\tderiveq \tderivtwo$ and $\sizetyp{\ltypetwo}=\sizetyp{\mset\ltypetwo} = \sizectx{\typctxtwo}$,
		so that $\insize{\tderivtwo} = 0 = \sizectx{\typctxtwo}- \sizetyp{\ltypetwo}$.

		\item \emph{Rule $\ruleAp$}, that is, $\tm = \neu \nf$. 
		Then $\tderiv$ has the following form:
\begin{center}$
		\tderiv = 
		\begin{prooftree}
		\hypo{\tderiv_{\neu} \derives \typctx_{\neu} \vdash \neu \hastype \tarrow{\mtype}{\ltype}}
		\hypo{\tderiv_{\nf} \derives \typctx_\nf \vdash \nf \hastype\mtype}
		\infer2[\footnotesize$\ruleApp$]{\tyjp{}{\neu \nf}{\typctx_{\neu} \mplus \typctx_\nf}{\ltype}}
		\end{prooftree}
$\end{center}
		where  $\typctx = \typctx_{\neu} \mplus \typctx_\nf$.
		By \ih (of Point 3), there exist a derivation $\tderivtwo_{\nf} \derives \typctxtwo_\nf \vdashun \nf \hastype\mtype'$ such that $\tderiv_\nf \tderiveq \tderivtwo_{\nf} $ and $\insize{\tderivtwo_\nf}=\sizectx{\typctxtwo_\nf} +\size{\mtype'}$. By \ih of Point 1 applied with respect to the linear type $\tarrow{\mtype'}\ltypetwo$, there exists a derivation $\tderivtwo_{\neu} \derives \typctxtwo_{\neu} \vdashun \neu \hastype \tarrow{\mtype'}\ltypetwo$ such that $\tderiv_\neu \tderiveq \tderivtwo_{\neu} $ and $\insize{\tderivtwo_\neu}=\sizectx{\typctxtwo_{\neu}} -\size{\tarrow{\mtype'}\ltypetwo} = \sizectx{\typctxtwo_{\neu}} -\size{\mtype'} - \size\ltypetwo -1$. Then the derivation $\tderivtwo$ of the statement is defined as follows:
\begin{center}$
		\tderivtwo \defeq 
		\begin{prooftree}
		\hypo{\tderivtwo_{\neu} \derives \typctxtwo_{\neu} \vdashun \neu \hastype \tarrow{\mtype'}{\ltypetwo}}
		\hypo{\tderivtwo_{\nf} \derives \typctxtwo_\nf \vdashun \nf \hastype\mtype'}
		\infer2[\footnotesize$\ruleApp$]{\typctxtwo_{\neu} \mplus \typctxtwo_\nf \vdashun \neu \nf \hastype \ltypetwo}
		\end{prooftree}
$\end{center}

		with $\typctxtwo \defeq \typctxtwo_{\neu} \mplus \typctxtwo_\nf$. Clearly $\tderiv \tderiveq \tderivtwo$. Moreover, 
		\begin{center}$\begin{array}{rcllllllllll}
		\insize{\tderivtwo} & = & \insize{\tderivtwo_{\neu}} + \insize{\tderivtwo_{\nf}} + 1 
		
		& =_{\ih} &\sizectx{\typctxtwo_{\neu}} - \sizetyp{\mtype'} - \sizetyp{\ltypetwo} -1+ \insize{\tderivtwo_{\nf}} + 1  
		\\
		& = &\sizectx{\typctxtwo_{\neu}} - \sizetyp{\mtype'} - \sizetyp{\ltypetwo}+ \insize{\tderivtwo_{\nf}}
		
		& =_{\ih} & \sizectx{\typctxtwo_{\neu}} - \sizetyp{\mtype'} - \sizetyp{\ltypetwo} + \sizectx{\typctxtwo_\nf} + \sizetyp{\mtype'}
		\\
		& = & \sizectx{\typctxtwo_{\neu}} - \sizetyp{\ltypetwo} + \sizectx{\typctxtwo_\nf}
		
		& = & \sizectx{\typctxtwo} - \sizetyp{\ltype}.
		\end{array}$\end{center}
\end{itemize}

\item \emph{$\tm$ is normal and $\type$ is a linear type $\ltype$}. Cases of $\tm$:
		\begin{itemize}
		\item $\tm$ is a neutral term, that is, $\tm=\neu$. Then by Point 1 with respect to $\ltypetwo = \tvar$ we obtain a derivation $\tderivtwo \derives \typctxtwo \vdashun \neu\hastype \tvar$ for which $\tderiv \tderiveq\tderivtwo$ and $\size\tderivtwo = \sizectx\typctx - \size\tvar= \sizectx\typctx = \sizectx\typctx + \size\tvar$. 

		\item \emph{$\tm$ is an abstraction}, that is, $\tm = \la{\var}{\nf}$ and  $\tderiv$ has necessarily the following shape:
		\begin{equation*}
		\begin{prooftree}
		\hypo{\tderiv'\derives \typctx' \vdash \nf \hastype \ltypetwo}
		
		\infer1[\footnotesize$\lambda$]{\typctx' \sm \var \vdash \la{\var}\nf \hastype \larrow{\typctx'(\var)}{\ltypetwo}}
\end{prooftree}
		\end{equation*}
		where $\ltype = \larrow{\typctx'(\var)}{\ltypetwo}$ and $\typctx = \typctx' \sm \var$.
		By \ih, there is a derivation $\tderivtwo' \derives \typctxtwo' \vdashun \nf \hastype \ltypethree$ such that $\tderivtwo' \tderiveq \tderiv'$ and $\insize{\tderivtwo'} = \sizectx{\typctxtwo'} + \size\ltypethree$. Then the derivation $\tderivtwo$ of the statement is defined as follows:
\begin{center}$\tderivtwo \defeq
		\begin{prooftree}
		\hypo{ \tderivtwo' \derives \typctxtwo' \vdashun \nf \hastype \ltypethree }	
		\infer1[\footnotesize$\lambda$]{\typctxtwo' \sm \var \vdashun \la\var\nf \hastype \larrow{\typctxtwo'(\var)}\ltypethree}
\end{prooftree}
$\end{center}
with $\typctxtwo \defeq \typctxtwo' \sm \var$ and $\typetwo \defeq \larrow{\typctxtwo'(\var)}\ltypethree$.
Clearly, $\tderiv \tderiveq \tderivtwo$. Moreover, 
		\begin{center}$
		\insize{\tderivtwo'} \ \  =_{\ih} \ \  \sizectx{\typctxtwo'} + \sizetyp{\ltypethree}
		\ \  = \ \  \sizectx{\typctxtwo'\sm\var} + \sizetyp{\typctxtwo'(\var)} + \sizetyp{\ltypethree}
		\ \  = \ \  \sizectx{\typctxtwo'\sm\var} + \sizetyp{\larrow{\typctxtwo'(\var)}{\ltypethree}}-1
		$\end{center}
		Therefore, 
		\begin{center}$\begin{array}{rcllllll}
		\insize{\tderivtwo} & = & \insize{\tderivtwo'} + 1
		
		& = & \sizectx{\typctxtwo'\sm\var} + \sizetyp{\larrow{\typctxtwo'(\var)}{\ltypethree}}-1 +1
		\\
		& = & \sizectx{\typctxtwo'\sm\var} + \sizetyp{\larrow{\typctxtwo'(\var)}{\ltypethree}}
		
		& = & \sizectx{\typctxtwo} + \sizetyp{\typetwo}.
		\end{array}$\end{center}
	\end{itemize}
	
	\item \emph{$\tm$ is normal and $\type$ is a multi type $\mtype$}. Then the last rule is necessarily $\ruleMany$. So, necessarily, for some finite set of indices $I$,
		\begin{equation*}
		\tderiv = 
		\begin{prooftree}
		\hypo{\tderiv_i\derives \tyjp{}{\tm}{\typctx_i}{\ltype_i}}
		\delims{\left[}{\right]_{\iI}}
		\infer1[\footnotesize$\ruleMany$]{\tyjp{}{\tm}{\mplus_{\iI}\typctx_i}{\mset{\ltype_i}_{\iI}}}
		\end{prooftree}
		\end{equation*}
		where $\typctx = \mplus_{\iI}\typctx_i$.
		By \ih (on $\tderiv_i$), there are derivations $\tderivtwo_i \derives \typctxtwo_i \vdashun \tm \hastype \ltypetwo_i$ such that $\tderivtwo_i \tderiveq \tderiv_i$ and $\insize{\tderivtwo_i} = \sizectx{\typctxtwo_i} + \size\ltypetwo_i$. The derivation $\tderivtwo$ of the statement is defined as follows:
		\begin{center}$
		\tderivtwo \defeq 
		\begin{prooftree}
		\hypo{\tderivtwo_i\derives \typctxtwo_i \vdashun \tm \hastype \ltypetwo_i}
		\delims{\left[}{\right]_{\iI}}
		\infer1[\footnotesize$\ruleMany$]{\mplus_{\iI}\typctxtwo_i \vdashun \tm \hastype \mset{\ltypetwo_i}_{\iI}}
		\end{prooftree}$\end{center}
		with $\typctxtwo \defeq \mplus_{\iI}\typctxtwo_i$ and $\typetwo \defeq \mset{\ltypetwo_i}_{\iI}$.
		Clearly, $\tderiv \tderiveq \tderivtwo$. Moreover, 		
		\begin{center}$\begin{array}{rcllllllllll}
		\insize{\tderivtwo} & = & \sum_{\iI}\insize{\tderivtwo_i} 
		
		& =_{\ih} & \sum_{\iI}\sizectx{\typctxtwo_i} +\sum_{\iI}\sizetyp{\ltypetwo_i}
		\\
		& = & \sizectx{\mplus_{\iI}\typctxtwo_i} +\sizetyp{\mset{\ltypetwo_i}_{\iI}}
		
		& = & \sizectx{\typctxtwo} + \sizetyp{\typetwo}.&\qedhere
		\end{array}$\end{center}
	\end{enumerate}
\end{proof}

%% file: 99-08-Head_Bounds_From_Composable_Types.tex
\section{Proofs of \Cref{sect:head-case} (The Less Satisfying Head Case)}
\label{app:head-case}

\begin{lemma}[Head normalization and composable types]
\label{lappendix:norm-and-comppairs-nonempty-head}
\NoteState{l:norm-and-comppairs-nonempty-head}
Let $\tm$ and $\tmtwo$ be closed terms. Then $\tm\tmtwo$ $\toh$-normalizes if and only if $\hcompairs\tm\tmtwo \neq \emptyset$. 
\end{lemma}

\input{\proofspath/head/normalization_composable_bounds}

\begin{theorem}[Lax bounds for head reduction from composable types]
	\label{thmappendix:lax-bounds-3-head}
	\NoteState{thm:lax-bounds-3-head}
Let $\nf$ and $\nftwo$ be closed normal terms such that $\deriv:\nf\nftwo \toh^* \hnf$ with $\hnf$ head normal. Then:
\begin{enumerate}
\item \emph{Lax bounds and types}:  $2\size{\deriv} + \hdsize{\hnf} \leq \size\ltype+\size\mtype+1$ for every composable pair $(\ltype, \mtype)\in \hcompairs\nf\nftwo$.
\item \emph{Lax bounds and types, up to substitutions}:  $2\size{\deriv} + \hdsize{\hnf} \leq \size\ltype+\size\mtype+1$ for every composable pair up to substitution $(\ltype, \mtype)\in \hsubcompairs\nf\nftwo$.
\end{enumerate}
\end{theorem}

\input{\proofspath/head/lax_composable_bounds}


%

%% file: proofs/head/normalization_composable_bounds.tex
\begin{proof}
If $\tm\tmtwo$ $\toh$-normalizes then by completeness (\refthm{head-characterization}) there exists a 
derivation $\tderiv$ for $\tm\tmtwo$. The last rule of $\tderiv$ is necessarily $\ruleAp$, and the premises of that rule 
provide a composable pair in $\hcompairs\tm\tmtwo$. Vice versa, if $\hcompairs\tm\tmtwo\neq\emptyset$ then every composable pair 
induces a derivation for $\tm\tmtwo$, by connecting the two derivations producing the composable pair via rule $\ruleAp$. Thus, $\tm\tmtwo$ is typable. By correctness (\refthm{head-characterization}), $\tm\tmtwo$ is $\toh$-normalizing.
\end{proof}

%% file: proofs/head/lax_composable_bounds.tex
\begin{proof} 
\hfill
\begin{enumerate}
\item Let $(\ltype, \mtype) \in \hcompairs\nf\nftwo$ be a composable pair, which implies $\ltype= \larrow{\mtype}{\ltypetwo}$ for some $\ltypetwo$. Then, there are two derivations $\concl{\tderiv_\nf}{}{\nf}{\larrow{\mtype}{\ltypetwo}}$ and $\concl{\tderiv_{\nftwo}}{}{\nftwo}{\mtype}$. We compose them via a $\ruleAp$ rule into a derivation $\tderiv$ for $\nf\nftwo$:
\begin{equation*}
		\tderiv \defeq 
		\begin{prooftree}
		\hypo{\tderiv_{\nf}\derives\vdash \nf \hastype \larrow{\mtype}{\ltypetwo}}
		\hypo{\tderiv_{\nftwo} \derives \vdash \nftwo \hastype\mtype}
		\infer2[\footnotesize$\ruleApp$]{\tyjp{}{\nf\nftwo}{}{\ltypetwo}}
		\end{prooftree}
		\end{equation*}
By head correctness (\refthm{head-characterization}), $2\size{\deriv} + \hdsize\hnf \leq \insize\tderiv = \insize{\tderiv_\nf} + \insize{\tderiv_{\nftwo}} + 1$. Now, since types bound the size of the derivation for normal terms (\refprop{types-bound-normal-derivations}),  we obtain $\insize{\tderiv_\nf} \leq \size{\larrow{\mtype}{\ltypetwo}}$ and $\insize{\tderiv_{\nftwo}} \leq \size\mtype$. Therefore,  $2\size{\deriv} + \hdsize\hnf \leq \insize\tderiv  \leq \size{\larrow{\mtype}{\ltypetwo}}+\size\mtype+1$, as required.	
\item Let $(\ltype, \mtype) \in \hsubcompairs\nf\nftwo$ be a composable pair up to substitution, which implies that there exist a type substitution $\sigma$ such that $\ltype\sigma= \larrow{\mtypetwo}{\ltypetwo}$, $\mtype\sigma = \mtypetwo$ for  some $\ltypetwo$ and $\mtypetwo$. Then, there are two  derivations $\concl{\tderiv_\nf}{}{\nf}{\ltype}$ and $\concl{\tderiv_{\nftwo}}{}{\nftwo}{\mtype}$. By stability of derivation under type substitution (\reflemmap{dry_is_standard}{sub}), applying $\sigma$ to $\tderiv_\nf$ and $\tderiv_{\nftwo}$ we obtain two derivations $\concl{\tderivtwo_\nf}{}{\nf}{\larrow{\mtypetwo}{\ltypetwo}}$ and $\concl{\tderivtwo_{\nftwo}}{}{\nftwo}{\mtypetwo}$ such that $\tderiv_\nf \tderiveq \tderivtwo_\nf$ and $\tderiv_{\nftwo} \tderiveq \tderivtwo_{\nftwo}$. We compose them via a $\ruleAp$ rule into a derivation $\tderivtwo$ for $\nf\nftwo$:
\[
		\tderivtwo \defeq 
		\begin{prooftree}
		\hypo{\tderivtwo_{\nf}\derives\vdash \nf \hastype \larrow{\mtypetwo}{\ltypetwo}}
		\hypo{\tderivtwo_{\nftwo} \derives \vdash \nftwo \hastype\mtypetwo}
		\infer2[\footnotesize$\ruleApp$]{\tyjp{}{\nf\nftwo}{}{\ltypetwo}}
		\end{prooftree}
\]
By head correctness (\refthm{head-characterization}), $2\size{\deriv} + \hdsize\hnf \leq \insize\tderivtwo = \insize{\tderivtwo_\nf} + \insize{\tderivtwo_{\nftwo}} + 1$. By the skeletal invariants (\reflemma{skel-equiv-properties}), we obtain $\insize{\tderivtwo_\nf} + \insize{\tderivtwo_{\nftwo}} + 1= \insize{\tderiv_\nf} + \insize{\tderiv_{\nftwo}} + 1$. Since types bound the size of the derivation for normal terms (\refprop{types-bound-normal-derivations}),  we obtain $\insize{\tderiv_\nf} \leq \size{\ltype}$ and $\insize{\tderiv_{\nftwo}} \leq \size\mtype$. Therefore,  $2\size{\deriv} + \hdsize\hnf \leq \size{\ltype}+\size\mtype+1$, as required.
\qedhere

\end{enumerate}
\end{proof}